\documentclass[aps,reprint,amsmath,amssymb,superscriptaddress]{revtex4-2}

\usepackage{amsmath}    
\usepackage{amssymb}
\usepackage{graphicx}   
\usepackage{verbatim}   
\usepackage[colorlinks,urlcolor=blue,citecolor=blue,linkcolor=blue]{hyperref}
\usepackage{color}      
\usepackage{subfigure}  
\usepackage{hyperref}   
\usepackage{dcolumn}
\usepackage{bm}
\usepackage{epsfig}
\usepackage{epstopdf}
\usepackage{braket}
\usepackage{bbold}
\usepackage{balance}

\begin{document}


\title{A quantum collisional classifier driven by information reservoir}

\author{Ufuk~Korkmaz}
\email[]{ufukkorkmaz@itu.edu.tr}
\author{Deniz T\"{u}rkpen\c{c}e}

\affiliation{Department of Electrical Engineering, \.{I}stanbul Technical University, 34469 \.{I}stanbul, Turkey}


\date{\today}

\begin{abstract}
We investigate the open dynamics of a probe qubit weakly interacting with distinct qubit environments bearing quantum information. We show that the proposed dissipative model yields a binary classification of the reservoir qubits' quantum information in the steady state in the Bloch qubit parameter space, depending on the coupling rates. To describe the dissipation model dynamics, we have adopted the collision model, in which the input information parameters of the reservoir qubits are easily determined. We develop a generalized classification rule based on the results of the micromaser-like master equation where the classification can be described in terms of the Bloch parameters. Moreover, we show that the proposed classification scheme can also be achieved through quantum parameter estimation. Finally, we demonstrate that the proposed dissipative classification scheme is suitable for gradient descent based supervised learning tasks.

\end{abstract}


\maketitle

\section{Introduction} \label{sec:introduction}
There has been growing interest in exploiting the advantages of quantum resources for machine learning (ML) and artificial intelligence (AI)~\cite{biamonte_quantum_2017,aimeur_quantum_2013,paparo_quantum_2014,dunjko_quantum-enhanced_2016,alvarez-rodriguez_supervised_2017,mitarai_quantum_2018}. Data classification is one of the basic subroutines of these fields of research. There are several proposals for quantum versions of perceptrons, the fundamental processing units of quantum neural networks (QNNs), as quantum classifiers relying on the existing advantages of the circuit model of quantum computation~\cite{schuld_simulating_2015,wan_quantum_2017,schuld_quantum_2018,tacchino_artificial_2019,torrontegui_unitary_2019,beer_training_2020}. 
However, in this model, the need for constructing unitary perceptron gates composed of multi-controlled operations deteriorates the algorithmic processes for existing quantum computers operating in the noisy intermediate-scale quantum (NISQ) era~\cite{preskill_quantum_2018}. Therefore, introducing alternative proposals for classifying quantum data with hardware-efficient protocols could be favourable. 

It has been reported that the dissipation-based computation model is equivalent to the gate-based quantum circuit model   ~\cite{verstraete_quantum_2009}. Alongside the fact that dissipation is detrimental to quantum information processing, it is now well-known that dissipation-assisted protocols could be harnessed as a resource for quantum computation tasks ~\cite{sinayskiy_efficiency_2012,schuld_quantum_2014,song_implementation_2015,marshall_classifying_2019}. Some studies report that quantum reservoirs may not necessarily be the trash dumps that the quantum information is thrown, however, they could be considered as communication channels where some information is transmitted to the system of interest~\cite{blume-kohout_simple_2005,zwolak_redundancy_2017}.   

We consider a probe qubit weakly interacting with distinct quantum information-bearing reservoirs. Here, the quantum information reservoirs~\cite{deffner_information_2013, deffner_information-driven_2013} are assumed to be strings of idealized quantum bits $\ket{\Psi(\theta,\phi)}$ with specific parameters. In analogy with quantum reservoir engineering~\cite{poyatos_quantum_1996}, the probe qubit is subjected to a dissipation process in the presence of information reservoirs where the steady-state is a non-trivial quantum state.  

Here, open system dynamics is described by a standard collision model~\cite{scarani_thermalizing_2002,ziman_diluting_2002,nagaj_quantum_2002} where the system sequentially contacts with identical, non-interacting environmental units in a finite time portion. This picture fits well with our definition of the information reservoirs, which provide the input data for classification. The repeated interaction-based collision model has become one of the successful approaches to describe the dynamics of various open quantum systems such as non-equilibrium systems~\cite{karevski_quantum_2009,seah_nonequilibrium_2019}, non-Markovian systems~\cite{ciccarello_collision-model-based_2013,kretschmer_collision_2016,mccloskey_non-markovianity_2014}, thermodynamics for quantum systems~\cite{turkpence_photonic_2017,seah_collisional_2019,de_chiara_quantum_2020} or quantum systems with strong coupling~\cite{strasberg_repeated_2019}.

In this study, we show how quantum data can be classified by dissipative quantum dynamics probed by a  quantum two-level system (TLS) in the weak coupling regime and discuss the suitability of a standard quantum collisional model to achieve this task. We introduce a general classification rule for the proposed model based on the analytical and numerical results.  We demonstrate that when a TLS interacts with distinct information environments, it responds with a binary decision in the steady-state depending on the reservoir parameters and the coupling rates.  It's also shown that collision models may provide an efficient route for dissipative quantum information processing tasks in addition to being an alternative approach to modelling open quantum dynamics.  

We explicitly demonstrate the functionality of the proposed model in performing the linear classification task examined by introducing the quantum noise in the parameter spaces to be classified. Moreover, we show that the classification task can be achieved by using quantum parameter estimation. Finally, we show that the dissipation-driven open quantum classifier is suitable for gradient-based supervised learning tasks as the function of parameters representing the behaviour of the training task has a smooth variation behaviour with proper differentiability.  

This manuscript is organized as follows. In Sec.~\ref{sec:preliminaries}, we give the essential definitions for the model we propose. In Sec.~\ref{sec:MDS}, we present the physical model and system dynamics with derived analytical expressions. In Sec.~\ref{sec:NA}, we present the numerical analysis of the proposed model with verifications of the analytical results. In Sec.~\ref{sec:CQFI}, we demonstrate binary classification via QFI. Then in Sec.~\ref{sec:OPL}, we apply our model to a supervised learning scheme with a gradient descent process. Finally, we conclude our study with the discussion section in Sec.~\ref{sec:conc}.

\section{Preliminaries}\label{sec:preliminaries}
\subsection{Classical model}
Perceptron is a mathematical model defining a binary classification task which is the simplest form of data classification~\cite{minsky_perceptrons_1987}. The model returns a binary output $z=f(\bf{x}^T \bf{w})$ modulated by an activation function $f$ where $\textbf{x}=[x_1,\ldots x_n]^T$ and $\textbf{w}=[w_1,\ldots w_n]^T \in \mathbb{R}^n$ are vectors corresponding to an input dataset and corresponding weights, respectively.  In this model, classification is performed as $z\equiv 0$ if $z=f(\bf{x}^T \bf{w}) \geq 0$ and $z\equiv 1$, else. Weighted summation of the data instances 
\begin{align}\label{Eq:Perceptron}
\textbf{x}^T \textbf{w}=\sum_i w_i x_i
\end{align}
are defined with adjustable weights $w_i$ in order to perform learning tasks.  A properly operating perceptron can linearly separate (see Fig. 1 (b)) the input data instances successfully. The activation function $f$ is, in general, a non-linear function in analogy with a biological neuron meeting the expressivity requirements. On the other hand, a single perceptron can still separate data instances with a linear (identity) activation function. However, activation functions are extremely important for multilayer neural networks as their structure affect the performance of learning tasks~\cite{apicella_survey_2021}. 

The proposed model concerns the linear classification of the quantum data with identity activation. However, the classifier model can be easily extended to an open quantum perceptron model by introducing a non-linear Hamiltonian~\cite{turkpence_reservoir_2020} in the collision process.  

\subsection{Quantum dissipative dynamics}
We resort to a dissipative model to describe the quantum classifier where the open quantum system $\varrho$ is defined by the weighted contribution of generators,
\begin{align}\label{Eq:Q-perceptron}
\frac{\partial\varrho}{\partial t}=P_1\mathcal{L}^{(1)}_t+\ldots +P_N\mathcal{L}^{(N)}_t
\end{align}
each representing non-unitary evolutions by distinct reservoirs. Here, in general, $\mathcal{L}^{(i)}_t$ are time-dependent and weighted by non-negative numbers $P_i$ representing the probability about which reservoir involves the evolution of the probe qubit $\varrho_0$. Eq.~(\ref{Eq:Q-perceptron}) is the quantum equivalent of Eq.~(\ref{Eq:Perceptron}) and represents a physical system only when the weak coupling condition is met, ensuring the validation of the additivity of the generators~\cite{filippov_divisibility_2017,kolodynski_adding_2018,mitchison_non-additive_2018}. 

In compliance with the repeated interactions protocol, the generators can be represented by completely positive trace-preserving (CPTP) maps    
\begin{align}\label{Eq:CPTP}
\Phi^{(i)}_t[\varrho_0]=\text{Tr}_{\mathcal{R}_i}\{U_t(\varrho_0\otimes\rho_{\mathcal{R}_i})U_t^{\dagger}\}.
\end{align}
In this case, $\rho_{\mathcal{R}_i}$ is the $i$th reservoir's quantum state, and $U_t$ is a unitary propagator that influences both the reservoir and the system. A dynamical map meeting $\Phi_{t+s}=\Phi_t(\Phi_s [\varrho])$ is referred to as a completely positive (CP) divisible map if it is CP for $t$ and $s\geq 0$. It has been demonstrated that, under the weak coupling condition, when cross-correlations between different reservoirs are prevented, CP divisibility guarantees the additivity of quantum dynamical maps~\cite{filippov_divisibility_2017,kolodynski_adding_2018}.    

With the above expressions in mind, the dynamics of the proposed model can be represented by the weighted combination of CP divisible maps given below that fulfil CP divisibility as
\begin{align}\label{Eq:CPTP addition}
\Phi_t[\varrho_0]=\sum_i P_i\Phi_t^{(i)}[\varrho_0]. 
\end{align}
Note that only if all the dynamical maps $\Phi_t^{(i)}$ in the summation are CP divisible, Eq.~(\ref{Eq:CPTP addition}) can be cast in place of Eq.~(\ref{Eq:Q-perceptron}). 

As previously stated, the binary classification result is encoded in the probe qubit in the steady state. In the Bloch representation, a qubit can be written as $\varrho=\frac{1}{2} ( \mathbb{1}+\boldsymbol\omega\cdot\boldsymbol\sigma )$, where $\boldsymbol\omega=(\omega_x,\omega_y,\omega_z)=(sin\theta cos\phi,sin\theta sin\phi, cos\theta)$ stands for the Bloch vector and $\boldsymbol\sigma=(\sigma_x,\sigma_y,\sigma_z)$ stands for a vector whose elements are the Pauli matrices.

As the focal point of our current study is to develop a dissipative classification process of quantum information over single qubit parameters, it's instructive to follow a quantum variant of the estimation theory. By using Fisher information, a random variable's unknown $\lambda$ parameter may be calculated. The following are the traditional Fisher details for a discrete random variable
\begin{align}\label{Eq:Classical Fisher}
\mathcal{F}_{\lambda}=\sum_r p_r(\lambda)\left[\frac{\partial \text{ln} p_r(\lambda)}{\partial\lambda}\right]^2
\end{align}  
where the probability of getting the outcome $r$ given the parameter $\lambda$ is denoted by $p_r (\lambda)$. The extension of Eq.~(\ref{Eq:Classical Fisher}) allows one to quantify quantum Fisher information (QFI) as 
\begin{align}
\mathcal{F}_{\lambda}=\text{Tr}[\rho_{\lambda} L^2_{\lambda}]=\text{Tr}[(\partial_{\lambda}\rho_{\lambda})L_{\lambda}] 
\end{align}  
where the symmetric logarithmic derivative, $L_{\lambda}$, is determined by $\partial_{\lambda} \rho_{\lambda}=\frac{1}{2}\{\rho_{\lambda},L_{\lambda}\}$
~\cite{helstrom_quantum_1969}. These formulations lead to the conclusion that, 
\begin{align}
\mathcal{F}_{\lambda}=&\sum_i \frac{(\partial_{\lambda}p_i)^2}{p_i}+\sum_i p_i 
\mathcal{F}_{\lambda,i}\\
&-\sum_{i\neq j}\frac{8 p_i p_j}{p_i+p_j}|\langle\psi_i|\partial_{\lambda}\psi_j\rangle|^2\nonumber
\end{align}
defines QFI for a generic mixed state $\rho_{\lambda}=\sum_i p_i\ket{\psi_i}\bra{\psi_i}$, where $\{ p_i\}$ are the eigenvalues of $\rho$ and $\mathcal{F}_{\lambda,i}$ is the QFI for a pure state with 
\begin{align}
\mathcal{F}_{\lambda,i}=4[\langle\partial_{\lambda}\psi_i|\partial_{\lambda}\psi_i\rangle-|\langle\psi_i|\partial_{\lambda}\psi_i\rangle|^2].
\end{align}
In addition, a convenient formula specific to TLS was developed as follows
~\cite{dittmann_explicit_1999,zhong_fisher_2013} 
\begin{align}\label{Eq:Fisher}
\mathcal{F}_{\lambda}=\text{Tr}[(\partial_{\lambda}\rho)^2]+\frac{1}{\text{det}\rho_{\lambda}}\text{Tr}[(\rho_{\lambda}\partial_{\lambda}\rho_{\lambda})^2].
\end{align}
Note that the density matrix $\rho_{\lambda}$, whose parameter is $\lambda$ to be estimated, will be considered in the steady state throughout our calculations. 
\section{Model and system dynamics}\label{sec:MDS}
Quantum collisional models are depicted by repeated interactions between an open quantum system and a sequence of ancilla units. The dynamics of the recurrent interactions are unitary with a small interaction time $\tau\rightarrow 0^+$ in the conventional quantum collisional model, and the ancilla units are identical and non-interacting. It is well known that the standard way of constructing collision models results in memoryless open quantum dynamics with equivalent master equations~\cite{filippov_divisibility_2017,ziman_all_2005}. 

\begin{figure}[!t]
\includegraphics[width=3.2in]{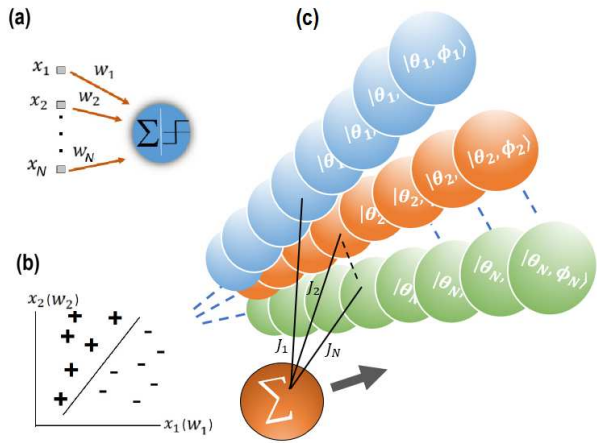}
\caption{ \label{fig:Fig1} (Colour online.) A description of the suggested model. A classical perceptron is seen in (a). Linear classification of a perceptron in the input-output space is shown in (b). The schematic depiction of our proposed quantum classifier is shown in (c).  } 
\end{figure}

As shown in Fig.~\ref{fig:Fig1} (c), the proposed model is described by the open quantum dynamics of a probe qubit repeatedly interacting with the units of $i=N$ distinct reservoirs. As mentioned in the previous section, we dub these reservoirs information reservoirs as each unit of the relevant reservoir is represented by pure, uncorrelated, identical qubit states   
\begin{align}\label{Eq:Inf Res}
\varrho_{\mathcal{R}_i}=\bigotimes_{k=1}^n\varrho_{i_k}(\theta_i,\phi_i).
\end{align} 
In order to express the additivity of quantum dynamical maps in terms of the quantum collisional model, each dynamical map $\Phi^{(i)}_t$ in Eq. (\ref{Eq:CPTP addition}) may be rephrased as
\begin{align}\label{Eq:Rephrase}  
\Phi^{(i)}_{n\tau}=&\text{Tr}_n \big[ \mathcal{U}_{0 i_n}\ldots\text{Tr}_1[\mathcal{U}_{0 i_1}\left(\varrho_0\otimes\varrho_{\mathcal{R}_{i_1}}\right)\mathcal{U}_{0 i_1}^{\dagger}]\otimes\ldots \nonumber \\ 
&\ldots\otimes\varrho_{\mathcal{R}_{i_n}}\mathcal{U}_{0 i_n}^{\dagger} \big]
\end{align}
by the reduced dynamics, where $n\tau$ is the period of time between $n$ collisions.

Here, $\mathcal{U}_{0 i_k}=\text{exp}[-\text{i}\mathcal{H}^k_{0 i}\tau]$ is the unitary propagator and $\mathcal{H}^k_{0 i}=\mathcal{H}^{k,i}_{\text{free}}+\mathcal{H}^{k,i}_{\text{int}}$ is the Hamiltonian describing the complete system in the course of interaction between the probe qubit and the $k$th ancilla of the $i$th reservoir where 
\begin{align}
\mathcal{H}^{k,i}_{\text{free}}=\frac{\hbar\omega_0}{2}\sigma_0^z+\frac{\hbar\omega_i}{2}\sum_{i=1}^N \sigma^z_{i_k}
\end{align}
stands for the free terms of the Hamiltonian; $\sigma_0^z$ is the Pauli-$z$ operator for the probe qubit, $\sigma^z_{i_k}$ is the Pauli-$z$ operator operating on the $k$th ancilla of the $i$th reservoir, and $\omega_{0,i}$ are the probe and reservoir qubit frequencies, respectively.

We consider $\omega_0=\omega_i$, in general, with different couplings such that
\begin{align}
\mathcal{H}^{k,i}_{\text{int}}=\hbar\sum_{i=1}^N J_i(\sigma_0^{+}\sigma_{i_k}^{-}+\text{H.c.})
\end{align}   
where $\sigma_{\nu}^{+}$ and $\sigma_{\nu}^{-}$ are the raising and lowering operators, respectively and $J_i$ is the coupling to the $i$th reservoir. We now omit the index $k$ for convenience. Here, the coupling strengths are in proportion $J_i \propto P_i$ to the probabilities corresponding to the generators, as expressed in Eq.~(\ref{Eq:Bloch}). 

To link the dynamical model of the research with actual physical systems, we suggest a master equation akin to a micromaser model based on repeated random interactions~\cite{cresser_quantum-field_1992, liao_single-particle_2010, turkpence_quantum_2016}. To that aim, we perform the system unitary evolution $\mathcal{U}(\tau)=\text{exp}[-\text{i}\mathcal{H}_{\text{int}}\tau]$ in the interaction picture with respect to $\mathcal{H}_{\text{free}}$ and evaluate the unitary operator (see Appendix~\ref{AppA})
\begin{align}\label{Eq:Unitary Expansion}
\mathcal{U}(\tau)=&\mathbb{1}-\text{i}\tau(\sigma_0^{+}\mathcal{S}^{-}_{j_i}+\sigma_0^{-}\mathcal{S}^{+}_{j_i}) \nonumber\\ 
&-\frac{\tau^2}{2}(\sigma_0^{+}\sigma_0^{-}\mathcal{S}^{-}_{j_i}\mathcal{S}^{+}_{j_i}+\sigma_0^{-}\sigma_0^{+}\mathcal{S}^{+}_{j_i}\mathcal{S}^{-}_{j_i})
\end{align}   
up to second order in $\tau$. Here, $\mathcal{S}^{\pm}_{j_i}=\sum_{i=1}^N J_i\sigma_i^{\pm}$ are the collective operators weighted by $J_i$. The whole system can be considered as factorized $\varrho(t)=\varrho_0(t)\otimes\varrho_{\mathcal{R}_i}$ after each interaction where reservoir states are assumed to be reset to their initial states. 

With reference to the micromaser theory ~\cite{filipowicz_theory_1986}, we use a Poisson process to describe the random interactions. Eq.~(\ref{Eq:Delta_t}) describes the dynamics of the system over a time interval $\delta t$, 
\begin{align} \label{Eq:Delta_t}
\varrho(t+\delta t)=r\delta t \mathcal{U}(\tau)\varrho(t)\mathcal{U}^{\dagger}(\tau)+(1-r\delta t)\varrho(t)
\end{align}
where $r\delta t$ is the likelihood of an interaction event occurring at a rate $r$ and $1-r\delta t$ is the likelihood that a non-interaction state would exist. One derives the following master equation for the probe qubit's reduced dynamics:
\begin{align}\label{Eq:MicroMaser}
\dot{\varrho}_0(t)=r\text{Tr}_{\mathcal{R}_i}[\mathcal{U}(\tau)\varrho(t)\mathcal{U}^{\dagger}(\tau)-\varrho(t)]
\end{align}
in the time limit $\delta t\rightarrow 0$ for $\dot{\varrho}_0(t)=(\varrho_0(t+\delta t)-\varrho_0(t))/\delta t$. 
The obtained master equation for the proposed model reads (see Appendix~\ref{AppB})
\begin{align}
\dot{\varrho}_0=&-i[\mathcal{H}_{\text{eff}},\varrho]+\sum_{i=1}^N J_i^2\left(\eta^{+}\mathcal{L}[\sigma_0^{+}]+\eta^{-}\mathcal{L}[\sigma_0^{-}]\right)
\nonumber\\
&+\sum_{i<j}^{N'} J_i J_j \left(\eta^{+}_s \mathcal{L}_s[\sigma_0^{-}]+\eta^{-}_s \mathcal{L}_s[\sigma_0^{+}]\right)
\end{align}
where $\mathcal{H}_{\text{eff}}=p\tau\sum_i^N J_i\left(\langle\sigma_i^{-}\rangle\sigma_0^{+}+\langle\sigma_i^{+}\rangle\sigma_0^{-}\right)$ denotes the effective Hamiltonian describing a coherent drive on the probe qubit. The averages calculated over identical reservoir units , i.e., $\langle \mathcal{O}_i\rangle=\text{Tr}[\mathcal{O}\varrho_{\mathcal{R}_i}]$. $\mathcal{L}[o]\equiv 2o\varrho o^{\dagger}-o^{\dagger}o\varrho-\varrho o^{\dagger}o$ is the 
Lindblad superoperator, while $\mathcal{L}_s[o]\equiv 2o\varrho o -o^2\varrho-\varrho o^2$  refers to the reservoir's squeezing effect. The coefficients of the Linbladians carry information corresponding to different entries of the reservoir units' density matrices. For instance, the standard Linbladian coefficients $\eta^{\pm}=p\tau^2 \langle \sigma_i^{\pm}\sigma_i^{\mp}\rangle/2$ contain diagonal entries of the $i$th information reservoir units, while $\eta_s^{\pm}=2p\tau^2 \langle \sigma_i^{\pm}\rangle\langle\sigma_j^{\pm}\rangle$ contains off-diagonal entries of the pairwise distinct reservoir units with $N'=N(N-1)/2$ terms in the summation.

Notice that the proposed classifier unveils the classification result at a steady state. The probe qubit at steady state is found to be    
\begin{align}\label{Eq:Steady}
\varrho_0^{\text{ss}}=&\frac{1}{J_{\sum}}\sum_{i=1}^N J_i^2\Big( \langle\sigma_{i}^+\sigma_{i}^-\rangle \ket{e}\bra{e} +\langle\sigma_{i}^-\sigma_{i}^+\rangle\ket{g}\bra{g}\nonumber\\
&+\left[ i\gamma_1^-\langle\sigma_z\rangle_i\ket{e}\bra{g}+\text{H.c.}\right]\Big)
\end{align}
where $J_{\sum}=\sum_i^N J_i^2$ and $\gamma^-_1=p\tau\sum_{i}^N J_i\langle\sigma_{i}^-\rangle.$
As apparent in the equation above, the steady state of the probe qubit yields non-vanishing off-diagonal terms manifesting that the information reservoirs are non-equilibrium environments. 

The input quantum information contained by the $i$th reservoir can be explicitly defined with the relevant parameters as
\begin{align}\label{Eq:RhoR}
\mathcal{\rho}_{\mathcal{R}_{i}}&=
\begin{bmatrix}
\frac{1+\cos\theta_{i}}{2} & \frac{e^{-i\phi_{i}}}{2}\sin \theta_{i} \\
\frac{e^{i\phi_{i}}}{2}\sin \theta_{i} & \frac{1-\cos\theta_{i}}{2}
\end{bmatrix}:=
\begin{bmatrix}
\langle\sigma_{i}^+\sigma_{i}^-\rangle & \langle\sigma_{i}^-\rangle \\
\langle\sigma_{i}^+\rangle & \langle\sigma_{i}^-\sigma_{i}^+\rangle
\end{bmatrix}.
\end{align}

We develop the classification rules through Pauli observables as the merit quantifiers of classification.
The non-equilibrium steady state nature of the model as expressed in Eq.~(\ref{Eq:Steady}) allows one to develop dissipative classification protocols also for qubit parameter $\phi$. We have reported that 
the binary decision rule for the azimuth qubit parameter $\theta$  can be read through steady state probe qubit magnetization as~\cite{korkmaz_transfer_2022}
\begin{equation}\label{BinaryCond2}
D^{\theta}:
\begin{cases}
class_1, & \langle\sigma_z\rangle_0^{ss} \geq 0        
\\
class_2, & \text{otherwise}.   
\end{cases}
\end{equation} 
where $\langle\sigma_z\rangle_0^{ss}=\frac{1}{J_{\sum}}\sum_i^N J_i^2 \cos\theta_i$ and $cos\theta_i:=\langle\sigma_z\rangle_i$. 
In the following sections, we analyse our analytical results with numerical verifications and develop a generalized binary rule expressing the classification of coherence through the qubit parameter $\phi$. The noise and the Poisson statistics effects will also be considered to examine the protocol's success.     

%

\section{Numerical Analysis}\label{sec:NA}
In this section, we analyze the system dynamics with numerical methods. The dynamics were depicted in Eqs.~(\ref{Eq:CPTP addition}) and (\ref{Eq:Rephrase}). We use the QuTip package \cite{johansson_qutip_2013} for numerical evaluation of the open system dynamics. Though we continue assuming that the reservoirs are ideal qubits, in realistic situations, noise effects will tend to deteriorate the classification performance of the probe qubit. Therefore to simulate the dynamics with realistic parameters, the dynamical generators in Eq.~(\ref{Eq:Rephrase}) can be redefined as
\begin{align}\label{Eq:Rephrase2}  
\Phi^{(i)}_{n\bar{t}}=&\text{Tr}_n \big[ \Lambda_{0 i_n}\ldots\text{Tr}_1[\Lambda_{0 i_1}\left(\varrho_0\otimes\varrho_{\mathcal{R}_{i_1}}\right)\Lambda_{0 i_1}^{\dagger}]\otimes\ldots \nonumber \\ 
&\ldots\otimes\varrho_{\mathcal{R}_{i_n}}\Lambda_{0 i_n}^{\dagger} \big]
\end{align}
where $\Lambda_{0 i}[\varrho]=-i[\mathcal{H}_{0i},\varrho]+\sum_{\nu}\mathcal{L}_0^{\nu}[\varrho]$ is a dynamical map satisfying CP divisibility. We notice that $\varrho=\varrho_0\otimes\varrho_{\mathcal{R}_i}$ and the Lindblad term $\mathcal{L}_0^{\nu}[\varrho]=\Gamma^{\nu}(2A_{\nu}\varrho A_{\nu}^{\dagger}-A_{\nu}^{\dagger}A_{\nu}\varrho-\varrho A_{\nu}^{\dagger}A_{\nu})$ acts on the probe qubit where $\nu=\lbrace \theta, \phi\rbrace$. Here, respectively, $\Gamma^{\theta}$ and $\Gamma^{\phi}$ energy dissipation and pure dephasing rates with $A^{\theta}=\sigma^{-}\otimes\mathbb{1}^{\otimes N}$ and $A^{\phi}=\sigma_z\otimes\mathbb{1}^{\otimes N}$. As is depicted below, we use $\bar{t}$ to express the elapsed time between successive interactions following the numerical methods. 

We follow the stochastic interactions process in our numerical calculations as in the derivation of the micromaser-like master equation. In this picture, among a large number of trials $K$, the number of successful interactions $k$ within a time interval is a fluctuating quantity. We can reasonably assume a binomial distribution for the random variable $k$ following 

\begin{equation}\label{Eq:Binomial}
b(k,K)=\binom{K}{k} p^k (1-p)^{K-k}
\end{equation}
where $0<p<1$ is the probability of occurring an interaction event with an expectation value $\langle k\rangle=\sum_k b(k,K)=pK$. 

The distribution in Eq.~(\ref{Eq:Binomial}) reaches the regular distribution for $p=1$ and reduces to the Poisson distribution for $p\rightarrow 0$, $K\rightarrow \infty$ so that $pK=\langle k\rangle$ is constant. 
Following these expressions, one defines the expectation of $k$ to be proportional with the average time interval $T$ as $\langle k\rangle=rT$ where $r$ is the interaction rate also defined in analytical calculations. Here, we define $\bar{t}=\overline{t_{k+1}-t_k}=1/r$ as the time between two successful interactions with overline being the average value. Working with realistic parameters conveniently, we choose $T=1/\Gamma^{\phi}$ so that $\langle k\rangle=r/\Gamma^{\phi}$ is the average number of successful interactions in the probe qubit lifetime. We define $\bar{t}=\tau+\tau_0$ as the sum of two different time scales where $\tau$ is the interaction time and $\tau_0$ is the time during a non-interaction event. Assuming $\tau$ is fixed for all interaction events and $\tau\ll 1/\Gamma^{\phi}$, we can approximate the dynamical map in Eq.~(\ref{Eq:Rephrase2}) as a unitary evolution during the interaction time $\tau$ and under the action of $\mathcal{L}_0^{\nu}[\varrho]$ during the interval $\tau_0$. 

\begin{figure*}
\includegraphics[width=7.3in]{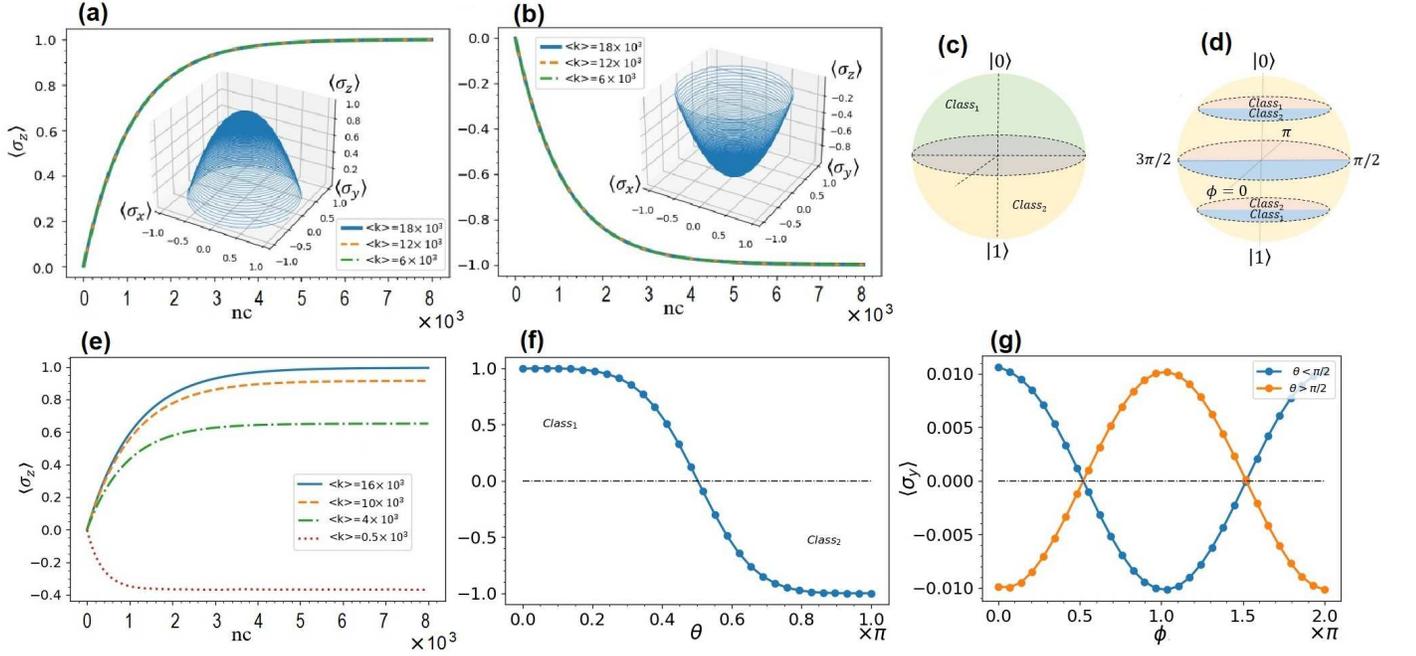}
\caption{ \label{fig:Fig2} (Colour online.) Temporal and input parameter-dependent steady state response of the probe qubit by a single information reservoir relative to the number of collisions (nc) based on the $\langle k \rangle$  mean interaction number. Identical reservoir units $\ket{\Psi(\theta,\phi)}$, (a) $\theta=0$, $\phi=0$ and (b) $\theta=\pi$, $\phi=0$ collided with the probe qubit prepared initially in the $\ket{+}=(\ket{e}+\ket{g})/\sqrt{2}$ state. For the case of multiple information reservoirs, the binary classification result is obtained according to which hemisphere of the Bloch sphere (c) the steady state magnetization value of the probe qubit will result. During the evolution, the Bloch vector trajectories with components  $\langle \sigma_{\nu}(n\bar{t})\rangle=\text{Tr}[\varrho_0(n\bar{t})\sigma_{\nu}]$ obtained for $\langle k \rangle=18\times 10^3$ interactions and depicted as the insets of (a) and (b). In (d), if there is more than one information reservoir, the binary classification result is obtained according to the $\phi$ amplitude parameter of the Bloch sphere, depending on the $\theta$ amplitude parameter in (c). During evolution, Bloch vector orbitals with components $\langle \sigma_{\nu}(n\bar{t})\rangle=\text{Tr}[\varrho_0(n\bar{t})\sigma_{\nu}]$ were obtained for $\langle k \rangle=18\times 10^3$ interactions and are shown as suffixes of (f) and (g). In (e), collision equilibration of noisy probe qubit magnetization by a single information reservoir delineated against the number of collisions (nc). This equilibration is  depending on the average number of random interactions in a $T$ time interval. $\theta=0$, is the amplitude parameter of the reservoir units. $\Gamma^{\theta}=2\times10^{-5}$, is the decay rate of the probe qubit. (f): The variation of the system qubit's steady-state magnetization. This variation is in the existence of a single information reservoir depending on $\theta$, with fixed $\phi=0$ interacting with the system qubit. In (g) steady-state magnetization variation of the system qubit interacting with a single information reservoir with different amplitude parameters:  blue line: $\theta=\pi/3$ fixed and depending on the $\phi$; orange line: $\theta = 2\pi/3$ fixed and depending on the $\phi$. $\tau=3$, the probe qubit-reservoir interaction time and $ J=0.01$, the coupling strength to the reservoir are dimensionless. They scale with $\omega_r$.} 
\end{figure*}

For the numerical calculations, we use the parameters of superconducting circuits\cite{blais_circuit_2021, krantz_quantum_2019,deng_robustness_2017}, which has become one of the most successful platforms for quantum information processing applications. Transmon qubits can be coupled via a resonator bus~\cite{majer_coupling_2007}, where the interactions are mediated by the exchange of virtual photons. In this architecture, the coupling strengths between qubits can be controlled via dispersive coupling to the transmission line resonator. A superconducting circuit with weakly coupled transmon qubits, typically, has a resonator frequency $\omega_r\sim 1-10$ GHz with $g\sim 1-500$ MHz qubit-resonator coupling and $J\sim 1-100$ MHz effective qubit-qubit coupling with qubit energy relaxation time $T_1\sim 40-150$ $\mu s $ and dephasing time $T_2\sim 50-100$ $\mu s $ \cite{majer_coupling_2007,deng_robustness_2017,research_ibm}. 

\subsection{Single information reservoir}

In Fig.~\ref{fig:Fig2}, we consider a single (only $J_1 \neq 0$) information reservoir $\varrho_1=\ket{\Psi(\theta_1,\phi_1)}\bra{\Psi(\theta_1,\phi_1)}$ in contact with a decay-free probe qubit. In this simplest case, the steady dynamics yield no classification result yet provides an instructive example of the so-called quantum homogenization. Here, homogenization amounts to an equilibration process in which the quantum state of the system becomes identical to that of the reservoir density matrix with diagonal entries~\cite{ziman_diluting_2002,nagaj_quantum_2002}.    

As we track the binary classification result through Bloch parameters, we monitor the system by Pauli observables. As shown in Figs.~\ref{fig:Fig2}~(a) and (b), the probe qubit magnetization converges to the magnetization of the reservoir units with the amplitude parameters $\theta_1=0$ and $\theta_1=\pi$, respectively. Aiming to provide a null magnetization initially, the probe qubit prepared as $\varrho_0=\ket{+}\bra{+}$.  The smooth and monotonic convergence of the equilibrium curves exhibits a Markov evolution demonstrating the success of CP divisible collision dynamics. The interaction statistics $\langle k\rangle$ does not affect the equilibration in the case of the idealized probe qubit.    
The insets of Figs.~\ref{fig:Fig2}~(a) and (b) depict the trajectory of the probe qubit Bloch vector during the dissipative process. In contrast to the equilibration curves, the Bloch vector exhibit an evolution with a smooth trajectory only for $\langle k \rangle > 18\times 10^3$ corresponding to regular interaction statistics. That would be significant for dissipative quantum processing tasks achieved by quantum collision dynamics.

As one may expect, the Bloch vector evolution ends up with the relevant hemisphere of the Bloch ball consistent with the quantum homogenization process. This trivial result will make sense when the equilibration of the probe qubit takes place in the presence of more than one information reservoir.  In this case, as depicted in Fig.~\ref{fig:Fig2}~(c) the equilibration dynamics yields the classification result depending on the probe qubit magnetization value.   

In what follows, we evaluate the collisional dynamics by introducing errors on the probe qubit, as depicted in Fig.~(\ref{fig:Fig2}) (e). For a smaller average number of successful interactions, the equilibration process breaks down as the average time $\bar{t}$ between successive interactions increases. Following the relation $\langle k\rangle=pK$, one could determine the $\langle k\rangle$ dependent probability of success $p$. Introducing the number of trials within $T$ as $K=T/\tau$, one can easily define $p=\langle k\rangle\tau/T$. For $p\simeq 1$, $p\simeq 0$ and $0<p<1$ one obtains regular, Poisson and sub-Poisson statistics respectively. 

In contrast to the cases where a single information reservoir state is fixed as in Figs.~\ref{fig:Fig2}~(a), (b) and (e), now in Figs.~\ref{fig:Fig2} (e) and (f), we evaluate the steady state response of the probe qubit to the variation of reservoir parameters $\theta$ (for fixed $\phi$ ) and $\phi$ (for fixed $\theta$), respectively. Both cases exhibit symmetric curves where we expect a similar response in the presence of more than one information reservoir due to the additivity of generators, as we found in Eq.~(\ref{Eq:Steady}). As a notable result in  Fig.~\ref{fig:Fig2} (g), the system has a reverse steady response curve against the variation of $\phi$, dependent on the fixed values of $\theta$.  

Therefore one may infer the parameter dependent classification rules from the steady system response in the presence of a single information reservoir as follows. First regarding the classification through the amplitude parameter, the binary decision of the system reads $Class_1$ for $\theta < \pi/2 $ and $Class_2$ otherwise. Second, the binary decision for $\phi$ reads $Class_1$ for $\pi/2 < \phi < 3\pi/2 $ and $Class_2$ otherwise. Note that this rule holds only for $\theta<\pi/2$  and the opposite holds for $\theta > \pi/2$. The results above, inferred from the single reservoir dynamics, will be examined for two reservoir cases in the following subsection and will aid in developing a generalized classification rule. 



\begin{figure}
\includegraphics[width=3.4in]{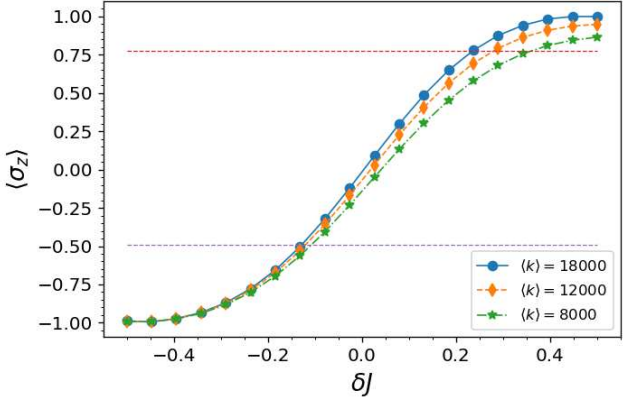}
\caption{ \label{fig:Fig3} (Colour online.) The probe qubit steady state response coupled to the two information reservoirs. Here, the system qubit; $J_1=J/2 + \delta J$ and $J_2=J/2 - \delta J$ are connected to the reservoirs by their coupling coefficients. $\delta J$ is a fraction of  $J$ with $J = 0.01$. $\langle k \rangle$, the average number of interactions within the $T=1/\Gamma^{\theta}$ time interval. With identical reservoir units of $\ket{\Psi(\theta_i,\phi_i)}$, $\theta_1=0$, $\phi_1=0$ and $\theta_2=\pi$, $\phi_2=0$, the probe qubit prepared initially in $\ket{+}=(\ket{e}+\ket{g})/\sqrt{2}$ state collisionally interacted. The longitudinal  decay rate of the probe qubit is $\Gamma^{\theta}=2\times10^{-5}$. Numerical and analytical verification was obtained for $0.774$ (red dashed line) and $-0.492$ (purple dashed line) values, respectively. Other parameters and the initial state of the probe qubit are the same as in the caption of Fig.~2. } 
\end{figure}

We observe that both regular and sub-Poisson interaction statistics yield successful equilibration dynamics needed for classification. Moreover, noise effects can be beaten with larger $\langle k\rangle$ as shown in Fig.~(\ref{fig:Fig3}). However, the current state-of-the-art for superconducting circuits set a limit for the possible minimum value for $\tau_0$.  
To successfully implement the collision dynamics, the probe qubit should be reset to its initial value to disentangle from the reservoir qubits after each interaction. This can be achieved by a qubit-reset protocol. 

Recently, it was stated that a parametric reset protocol could be performed at 34 ns in transmon qubits and could be reduced to 10 ns in principle~\cite{zhou_rapid_2021}. In our dissipative classification scheme with a noisy probe qubit, ~30 ns qubit reset time (including the qubit preparation time for the next collision) corresponds to $\langle k \rangle = 1/\Gamma_0(\tau_0+\tau)\simeq 12\times 10^3$ successful collisions. 

\begin{figure*}
\includegraphics[width=7.3in]{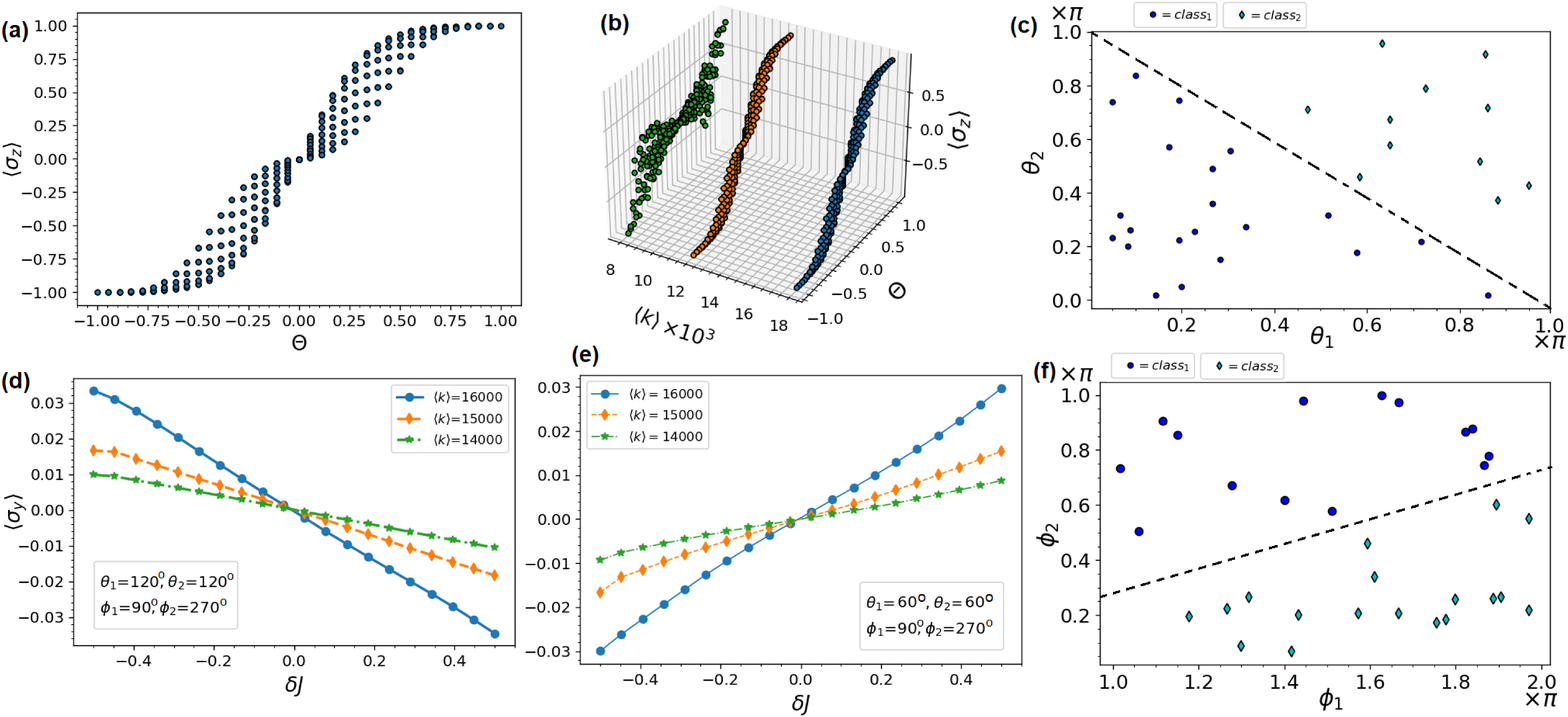}
\caption{ \label{fig:Fig4} (Colour online.) The system's steady-state response as a function of the linear change of the input parameters. A change in the steady state magnetization of a system qubit connected to various surroundings that convey information and are parametrized by $\theta$ for (a) and (b). When the $\theta = 0^{\circ}, 10^{\circ} \ldots 180^{\circ}$ azimuthal angles are present, $19 \times 19 = 361$ dots are presented, each representing steady state magnetization in the first and second environments. System to reservoirs coupling are fixed and equal to $J_1=J_2 = 0.01$. In both (a) and (b), the magnetization is plotted versus $\Theta= \pi-(\theta_1 + \theta_2 )$ for practical scaling as described in the text. (a) The change in the steady state magnetization of the system qubit as a function of the average number of interactions $\langle k \rangle$ (16000) throughout a time interval $T=1/\Gamma^{\phi}$. (b) How the steady state magnetization of the system qubit varies based on three distinct average interaction values of $\langle k \rangle$. (c) Linearly separated binary classification pattern of the system qubit in the steady state coupled to two information reservoirs with equal strengths. The pattern is obtained against the variation of $\theta$ parameters in range $0< \theta_{1,2} < \pi$, with the randomly chosen $32$ amplitude parameter pairs. For (a), (b) and (c) the remaining parameters and the initial state of the probe qubit are the same as in the caption of Fig.~2. For (d) and (e), the system qubit's steady state magnetization varies according to the coupling coefficients $J_1=J/2 + \delta J$ and $J_2=J/2 - \delta J$, where $\delta J$ is a portion of $J$ with $J = 0.01$. The average number of interactions $\langle k \rangle$ therein the times $T_1=1/\Gamma^{\theta}$ and $T_2=1/\Gamma^{\phi}$. With identical reservoir units of $\ket{\Psi(\theta_i,\phi_i)}$ with for (d) $\theta_1=\theta_2=2\pi/3$, $\phi_1=\pi/2$, $\phi_2=3\pi/2$ and for (e) $\theta_1=\theta_2=\pi/3$, $\phi_1=\pi/2$, the probe qubit prepared initially in $\ket{\Psi(\theta,\phi)}$ with $\theta=0$, $\phi=0$ state collisionally interacted. The energy dissipation rate of the probe qubit is $\Gamma^{\theta}=9.09\times10^{-6}$ and the dephasing time is $\Gamma^{\phi}=7.41\times10^{-6}$. (f) The system qubit's linearly separable classification pattern in steady state. The couplings between the information reservoirs and the system qubit were assumed to be equivalent. $32$ amplitude parameter pairs were classified randomly in the ranges of $0< \phi_{1} < \pi$,~$\pi < \phi_{2} < 2\pi$ and $\theta_1=\theta_2=\pi/3$. The interaction rate $r$ and the interaction duration $tau$ are assumed as $r=0.16$, $tau=3.00$, respectively, for (d), (e), and (f). } 
\end{figure*}

\subsection{Two information reservoirs}
Now, we consider the simplest case in which the concept of classification can be mentioned, where two information reservoirs interact with the probe qubit. In this case the respective couplings are $J_1, J_2 \neq 0$ and the information reservoirs are, in general characterized by $\varrho_{1,2}=\ket{\Psi(\theta_{1,2},\phi_{1,2})}\bra{\Psi(\theta_{1,2},\phi_{1,2})}$ specific parameters. As the first numerical demonstration of a multi-information reservoir case, we analyse the steady response of the noisy probe qubit with the variation of the coupling strengths. In Fig.~\ref{fig:Fig3}, one can see a slight effect of interaction statistics on the symmetric response curve. The appearance of the symmetrical curve as in Fig.~\ref{fig:Fig2} (f), where the dynamics were analysed in the presence of a single reservoir, manifests the adequacy of the model for binary classification. 

In case of two reservoirs, the steady response for the classification due to parameter $\theta$ can be given by (see Appendix~\ref{AppB}) the Pauli observable $\langle \sigma_z^0 \rangle^{ss}= \frac{1}{J_{\sum}}\times ( J_1^2 \cos \theta_1+J_2^2 \cos \theta_2)$ where 
$J_{\sum}=J_1^2+J_2^2$. For instance, when two information reservoirs (with $\theta_1=0, \phi_1=0$ and $\theta_2=\pi, \phi_2=0$) collisionally couple to the probe qubit with arbitrary coupling rate pairs, one obtains $\langle \sigma_z^0 \rangle^{ss}(J_1=0.0737,J_2=0.0026)=0.0774$ and $\langle \sigma_z^0 \rangle^{ss}(J_1=0.0036, J_2=0.0631)=-0.492$. These results obtained by analytical expressions, are represented by horizontal lines in Fig.~\ref{fig:Fig3} and show that they agree with numerical results corresponding to regular statistics. 

In addition to the above-described situation where the reservoir parameters are fixed and the coupling constants vary, we have also examined the system response according to the situation where the couplings are fixed and the reservoir parameters vary. First, we consider two reservoirs with different azimuthal Bloch parameters $\theta_1$ and $\theta_2$. The plotted dots are steady states with equal and fixed couplings. In particular, we take $\theta_1=0^{o}$ in degrees and a set of 19 parameter pairs  $\{\theta_1=0^{o};\theta_2=0^{o},\cdots ,\theta_1=0^{o};\theta_2=180^{o}\} $ were considered where $\theta_1$ kept fixed. Next, another 19 parameter pairs plotted with fixed $\theta_1=10^{o}$ and totally the steady response of 361 azimuthal Bloch parameter pairs numerically calculated (see Fig. \ref{fig:Fig4} (a) ). Here, we note that the pattern obtained in Fig.~\ref{fig:Fig4} (a), belongs to average $\langle k \rangle=18\times 10^3$ interactions very close to the case of the regular interactions.  The effect of statistics was also considered in  Fig.~\ref{fig:Fig4} (b) with a different number of average interactions $\langle k \rangle$ where quantum noise significantly involved the dynamics. As depicted in Fig.~\ref{fig:Fig1} (b), the success of the proposed classifier can be best demonstrated by the linear classification pattern of the input quantum data instances. Therefore, we examine the
randomly generated input data pairs parametrized by $\theta$ with the classification rule in Eq.~\ref{BinaryCond2}. Fig.~\ref{fig:Fig4} (c) demonstrates the operational classification success of the proposed model in the $\theta$ parameter space.  
    
As mentioned in the previous sections, our model also proposes a classification scheme in terms of the Bloch parameter $\phi$. To examine this scheme numerically, we obtained the system response through the Pauli observable $\sigma_y$ in the steady state. Figs.~\ref{fig:Fig4} (d), (e) exemplifies the suitability of classifying the input quantum information through the parameter $\phi$ in the presence of two information reservoirs with fixed parameters. Note that, the parameter $\theta$ of the two reservoirs is set equal to examine the classification over the parameter $\phi$.  
Following Eq.~(\ref{Eq:Rephrase2}), the dynamics are examined again by considering the noise effects. The consideration of the steady-state dynamics according to the variation of the $\delta J$ parameter determines which of the two reservoirs whose only $\phi$ parameters differ will weigh the steady-state system response. We have the curves obtained as the steady response symmetric relative to $\langle \sigma_y\rangle =0$ axis indicating that the proposed model can successfully achieve the binary classification through $\phi$. 

Here, the notable point is that the classification rule changes sign when the $\theta$ parameters of the two reservoirs belong to a Bloch vector in the different hemispheres of the Bloch ball. This result agrees with the numerical result in Fig.~\ref{fig:Fig2} (g), where we examine the single qubit reservoir case. As a more generic conclusion, one can deduce that the binary classification result through $\phi$  depends on the binary classification result of the parameter $\theta$. Therefore, the binary decision ($D^{\phi}$) for parameter $\phi$ reads
\begin{equation}\label{BinaryCond3}
(D^{\phi}|\langle\sigma_z\rangle_0^{ss}\geq (<)~0):
\begin{cases}
class_{1(2)}, & \langle\sigma_y\rangle_0^{ss} \geq (<)~0        
\\
class_{2(1)}, & \text{otherwise}.   
\end{cases}
\end{equation}   
Together with Eq.~(\ref{BinaryCond2}), the expression above introduces a general classification rule for a dissipative quantum classifier driven by quantum information environments. Fig.~\ref{fig:Fig4} (f) depicts the success of the proposed classifier in the two-dimensional $\phi$ parameter space. 

\begin{figure}[!t]
\includegraphics[width=3.2in]{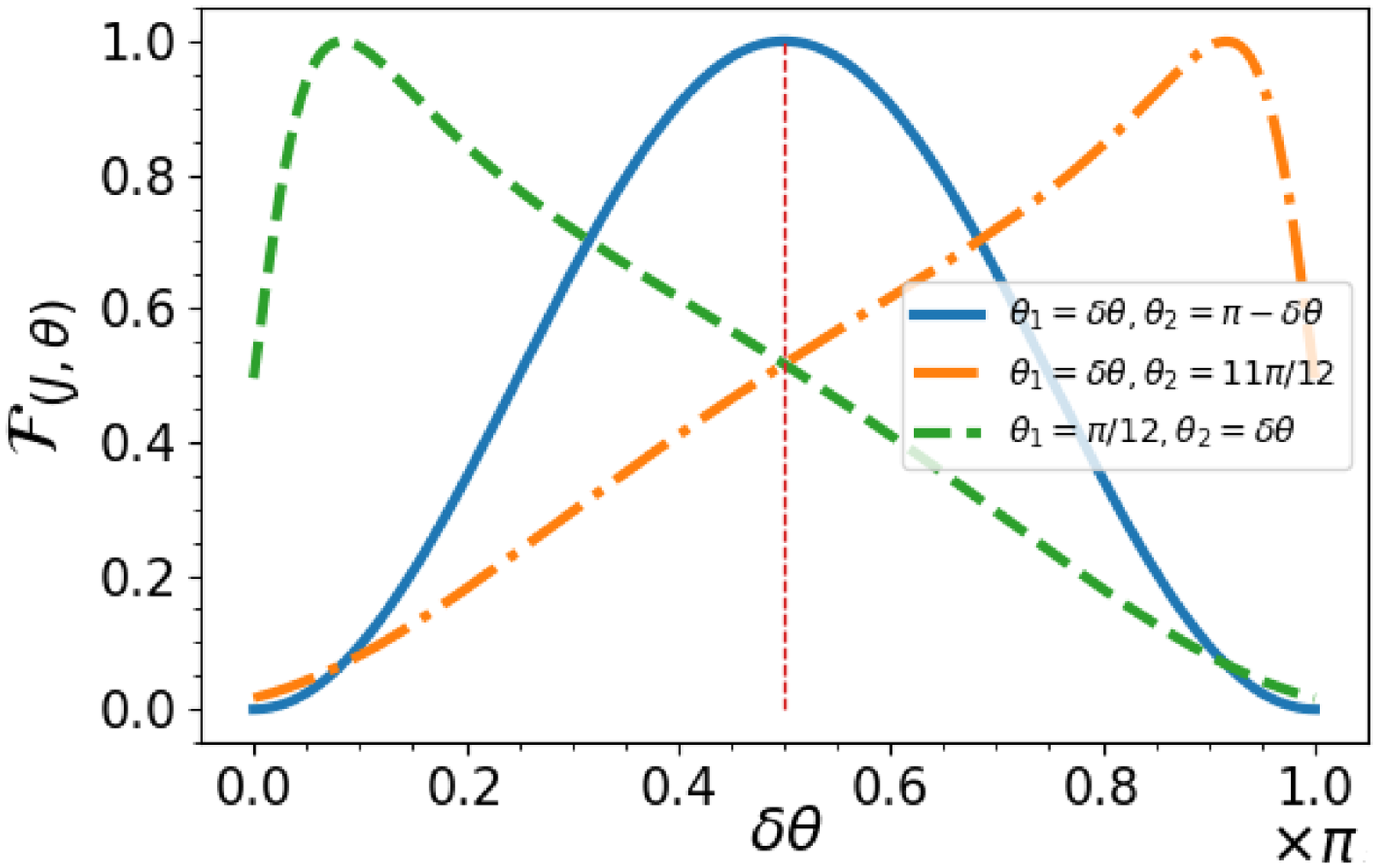}
\caption{\label{fig:Fig5}(Colour online.)The steady state probe qubit's QFI depends on three distinct amplitude parameter pairs -$\theta_1$,$\theta_2$- in the presence of two information reservoirs. QFI calculated against $0\leq\delta\theta <\pi$. $\theta_2=11\pi/12$ for the amplitude parameter of the second reservoir and $\theta_1=\pi/12$ for the amplitude parameter of the first reservoir are fixed values that correspond to the second and third situations, respectively. Based on $\delta\theta$, as shown in the legend, the other parameters are determined. The probe qubit's coupling to both reservoirs are configured to be equal and $J1=J2=0.01$. The interaction rate is assumed to be $r=0.20$ and the interaction duration is assumed to be $\tau=3.00$.} 
\end{figure}

\section{Classification through QFI}\label{sec:CQFI}
As described above, the binary decision of the proposed model is non-unitarily encoded on the probe qubit. 
The use of a probe connected to an environment is similar to the studies concerning quantum sensing~\cite{tan_quantum_2022,sha_continuous-variable_2022,wu_non-markovian_2021}. Here, QFI is the tool to estimate an unknown parameter. Although our parameters are well defined and can be monitored by certain observables, in this section, we show that binary classification can also be achieved through QFI. By Eq.~(\ref{Eq:Fisher}), one obtains the QFI expression for any parameter $\lambda=\lbrace\theta,\phi\rbrace$. The probe qubit density matrix in the steady state is explicitly defined in Eq.~(\ref{Eq:STTP}) as a function of $\lambda$.  

Following the recipe above, the QFI for the respective parameters in the presence of a single information reservoir is found to be 

\begin{figure}[!t]
\includegraphics[width=3.2in]{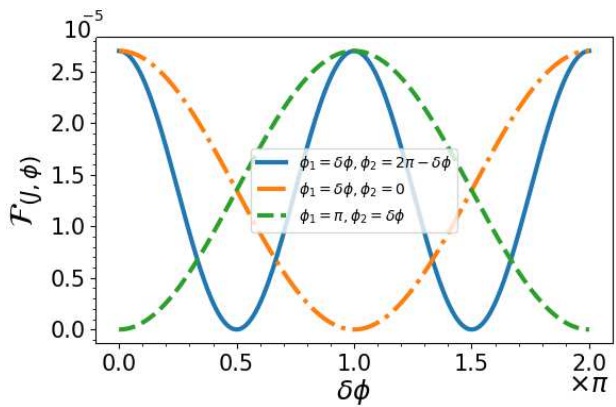}
\caption{\label{fig:Fig6}(Colour online.) The steady state probe qubit's QFI depends on three distinct amplitude parameter pairs -$\phi_1$,$\phi_2$- in the presence of two information reservoirs. Here, $\theta_1=\theta_2=\pi/6$ amplitude parameter pairs are set to match to the other situations. QFI calculated against $0\leq\delta\phi <2\pi$. $\phi_2=0.00$ for the amplitude parameter of the second reservoir and $\phi_1=\pi$ for the amplitude parameter of the first reservoir are fixed values that correspond to the second and third situations, respectively. Based on $\delta\phi$, as shown in the legend, the other parameters are determined. The probe qubit's coupling to both reservoirs are configured to be equal and $J1=J2=0.01$. The interaction rate is assumed to be $r=0.20$ and the interaction duration is assumed to be $\tau=3.00$.} 
\end{figure}

\begin{align}\label{Eq:QFIT}
\mathcal{F}_{\theta}=&\frac{\sin^2\theta}{2}+2\xi^2\cos^22\theta+\frac{1}{\sin^2\theta-\xi^2\sin^22\theta}\nonumber\\
&\Big(\frac{\sin^2\theta}{2}+\frac{\sin^22\theta}{8}+\xi^2(\frac{-3}{2}\sin2\theta\sin4\theta\nonumber\\
&+\cos^22\theta+\cos^2\theta\cos^22\theta-\sin^2\theta\sin^22\theta\nonumber\\
&+\frac{\xi^2}{2}\sin^24\theta)\Big) 
\end{align}
\begin{align}\label{Eq:QFIP}
\mathcal{F}_{\phi}=&\xi^2\sin^22\theta 
\end{align}
where $\xi=\tau r J/2$. Note that in the single reservoir case, as expressed in Eq.~(\ref{Eq:QFIP}), $\mathcal{F}_{\phi}$ only depends on $\theta$. However, the dependence to $\phi$ appears in the general expression as in Eq.~(\ref{Eq:FisherPhi}). General QFI expression for $\theta$ is too cumbersome to calculate analytically. Therefore, we evaluate the two information reservoir case numerically in Fig.~\ref{fig:Fig5} by use of Eq.~(\ref{Eq:DPHI}). Here, the system couplings to both the reservoirs are equal and three different values of reservoir parameter pairs $\theta_1$, $\theta_2$ (with $\phi_i=0$) are examined. First, $\theta_1=\delta\theta$ and $\theta_2=\pi-\delta\theta$ are taken and as $\delta\theta$ varies both parameters become equal when $\theta_1=\theta_2=\pi/2$. In this case, the QFI of the probe qubit reaches a maximum value (solid line). One obtains the same curve in the presence of a single information reservoir. Second, $\theta_2=11\pi/12$ is fixed and the QFI is evaluated under the variation of $\delta\theta$ where QFI, again, reaches the maximum value (orange dashed-dotted line) when both reservoir parameters are $\theta=11\pi/12$ equal. When the amplitude parameter $\delta\theta>\pi/2$ corresponds with the right side of the $\delta\theta=\pi/2$ axis, the probe qubit's magnetization will point within the southern hemisphere of the Bloch sphere (matching to classification label `1'). Similar justifications allow us to analyze the classification outcomes in the third instance, where the first reservoir parameter is set at $\theta_1=\pi/12$, with the binary label `0'. As a result, the binary decision of the model is determined by whether the QFI of the probe qubit falls in the right or left area of the vertical dashed line.

In a real parameter estimation scheme, the input (reservoir) parameters are fixed. Therefore, a trivial input with adjustable parameters is needed to monitor the maximum values of the QFI. Similar to the situation in Fig.~\ref{fig:Fig5}, binary classification may be done using the $\text{max}\{\mathcal{F}(\delta\theta_t)\}$ corresponding value of the trial reservoir amplitude parameter $\delta\theta_t$. Given the aforementioned expressions, the binary classification for QFI is as follows:
\begin{equation}\label{BinaryCond3a}
Decision:
\begin{cases}
0, & \delta\theta_t\geq\pi/2 \triangleq  \text{max}\{\mathcal{F}(\delta\theta_t)\} 
\\
1, & else. 
\end{cases}
\end{equation}

Eq.~(\ref{Eq:FisherPhi}) is the analytically obtained QFI for $\phi$. As shown in Fig~\ref{fig:Fig6}, we assess QFI for two information reservoirs. In this section, we take into account three scenarios in which the amplitude parameters of two information reservoirs are specified. Keep in mind that for simplicity's sake, the couplings of the probe qubit to both reservoirs are equal. First we consider, $\phi_1=\delta\phi$ and $\phi_2=2\pi-\delta\phi$. In this case, as we change $\delta\phi$, both the azimuth parameters reach the same value $\phi_{1,2}=0$ at $\delta\phi=0,\pi,2\pi$, which maximizes QFI (solid line). 
Second, $\phi_2=0$  is a fixed value, and we assess QFI under the influence of $\delta\phi$, where QFI takes the highest value when both reservoir parameters are identical at $\phi_{1,2}=0$ at $\delta\phi=0,2\pi$. In the last scenario, $\phi_1=\pi$ is constant, and we assess QFI while varying $\delta\phi$, with QFI once more taking the highest value when both reservoir parameters are $\phi_{1,2}=\pi$ identical.

\section{Open quantum learning}\label{sec:OPL}
Supervised learning involves the processes of iterative minimization of the cost function between an actual and a target output. A system parameter (in classic schemes the weights) is updated during the iteration for cost function minimization. In order to adapt the problem to our case, we choose to update the coupling strengths of the reservoirs to the probe qubit as
\begin{equation}\label{Eq:training}
\bf{J}_{k+1}=\bf{J}_{k}+\delta\bf{J}_{k}.
\end{equation}
where $\delta\bf{j}$ denotes a small change in $\bf{j}$. In an analogy to least squares approaches, the cost function is defined as the square difference between the vectors representing the actual $\bf{A}$ and the desired values $\bf{Y}$ as
\begin{equation}\label{Eq:Cost1}
C=\frac{1}{2}(\bf{Y}-\bf{A})^2.
\end{equation}
In our problem, the cost function reads
\begin{equation}\label{Eq:Cost2}
C=\frac{1}{2}(\langle \sigma_z^0\rangle_{des}^{ss}-\langle \sigma_z^0\rangle_{act}^{ss})^2
\end{equation}
where the $\langle \sigma_z^0\rangle_{des}^{ss}$ desired and the $\langle \sigma_z^0\rangle_{act}^{ss}$ actual steady state magnetizations of the proposed classifier. 
\begin{figure}[!t]
\includegraphics[width=3.2in]{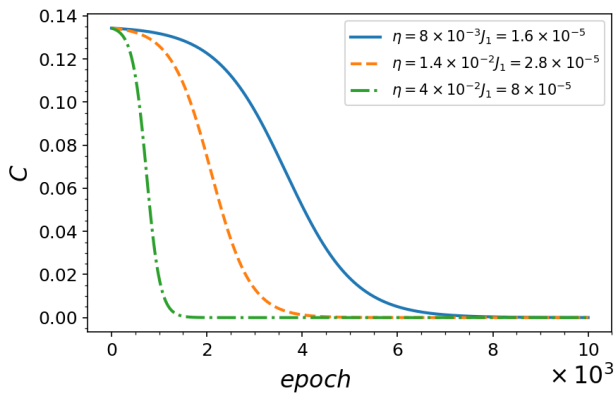}
\caption{\label{fig:Fig7}(Colour online.)  GD outcomes based on three distinct learning rates' coupling coefficients. Here, the initial values  $\langle \sigma_z^1\rangle=0.94$, $\langle \sigma_z^2\rangle=-0.10$, $g_1=0.002$, $g_2=0.05$ and $\langle \sigma_z^0\rangle_{des}^{ss}=0.42$, respectively.}  
\end{figure}
The well-known gradient descent method is used in the protocol to reduce costs~\cite{wan_quantum_2017}. The parameters are iterated by
\begin{equation}\label{Eq:delta_nu}
\delta {j}_i=-\eta\frac{\partial C}{\partial {j}_i}.
\end{equation}
where $\eta$ is the so-called learning rate defining the speed of the variation process in the greatest decrease direction. 

\begin{figure}[!t]
\includegraphics[width=3.2in]{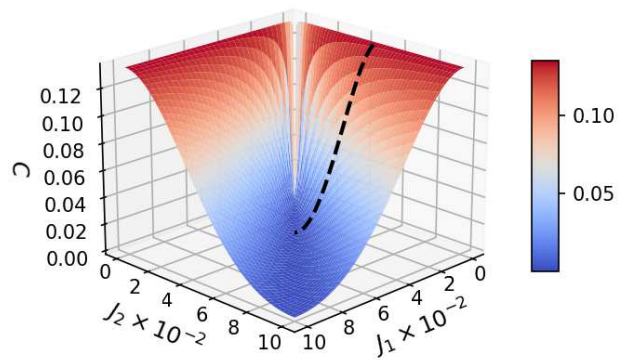}
\caption{\label{fig:Fig8}(Colour online.) The cost function's 3D surface plot according to $J_1$ and $J_2$. Here, the initial values for the dashed line are the learning rate $\eta=2.6\times10^{-5}$, $\langle \sigma_z^1\rangle=0.94$, $\langle \sigma_z^2\rangle=-0.10$, $J_1=0.002$, $J_2=0.05$ and $\langle \sigma_z^0\rangle_{des}^{ss}=0.42$, respectively.}  
\end{figure}

We used the GD approach to train our model for three different learning rates, as shown in Fig.~(\ref{fig:Fig7}) using Eq.~(\ref{Eq:training_1}). The smooth variation of the cost function against the iteration of the coupling rates exhibits a successful descent of the model, even for high learning rates with avoided overshooting. Additionally, as shown in Fig.~(\ref{fig:Fig8}), our model offers a good 3D graphical representation of the cost function. The cost function is shown in Fig.~(\ref{fig:Fig8}) as decreasing along dotted lines when $J_1$ and $J_2$ parameter changes are successful.

\section{CONCLUSIONS}\label{sec:conc}
In conclusion, we have explored the dynamical behaviour of a single TLS repeatedly interacting with structured environments parametrized by the Bloch variables. We observe that by monitoring the TLS, acting as a probe system by Pauli observables, one obtains a binary decision in the steady state as the classification result of $N$ distinct information reservoirs as input data. 

Our preference for a dissipative protocol for quantum data classification allows one to obtain the result as an open system response of a natural equilibration process in contrast to the circuit-based unitary schemes where the number of multi-qubit gates scales exponentially with the number of input data. Furthermore, our preference for a collision model to characterize the dissipative dynamics allows us to easily parameterize the information reservoir units as input data.  

We obtained an analytical statement for the proposed model by deriving a micro-maser master equation as a physical example expressing repeated interactions. Based on the analytical expressions, we derived a generalized classification rule for the Bloch parameters' space. We also examined the analytical results with numerical verifications introducing noise with realistic parameters affecting both the Bloch parameters. We have demonstrated that the model is feasible by the current state-of-the-art. Moreover, the classification rule based on the dissipative dynamics is considered upon quantum parameter estimation and introduced a classification rule through QFI. 

Finally, we apply our model to a supervised learning scheme with a gradient-descent process as an ultimate goal. We derived the cost function straightforwardly and observed the successful minimization by dissipative dynamics. The proposed model can be considered as a unit structure of a dissipation-driven, feed-forward quantum network once a non-linear function is introduced at a certain threshold during the collisional dynamics. Such a scheme can be favourable when applied in place of classification gates in the circuit model variational quantum algorithms to minimize errors in the current NISQ devices known as hybrid quantum computing. 

\begin{acknowledgments}
The authors gratefully thank funding from Turkey's Scientific and Technological Research Council (TÜBTAK-Grant No. 120F353). The authors would also like to thank the Cognitive Systems Laboratory in the Department of Electrical Engineering for providing a supportive atmosphere for the research.
\end{acknowledgments} 
  
\appendix
\section{The Unitary Propagator}\label{AppA}

The unitary time-evolution operator is written as $\mathcal{U}(\tau)=\text{exp}[-\text{i}\mathcal{H}_{\text{int}}\tau]=\text{exp}[-i\tau U]$. Here, we denote 
\begin{align}\label{Eq:U_Sum}
U=\sum_{i=1}^N J_i(\sigma_0^{+}\sigma_i^{-}+\text{H.c.}). 
\end{align}
by taking $\hbar=1$. We make an approximation of the second-order unitary propagator in $\tau$. 
\begin{equation}\label{Eq:U_Approx}
\mathcal{U}(\tau)\simeq \mathbb{1}-i\tau U-\frac{\tau^2 U^2}{2}
\end{equation}
The matrices for the operators $U$ and $U^2$ are as follows:
\begin{subequations}
\begin{align}
& U=
\begin{bmatrix}\label{Eq:U}
 & \mathcal{S}^{-}_{j_i}\\
\mathcal{S}^{+}_{j_i} & 
\end{bmatrix}\\
& U^2=
\begin{bmatrix}\label{Eq:U_Sq}
 \mathcal{S}^{-}_{j_i}\mathcal{S}^{-}_{j_i} & \\
 & \mathcal{S}^{+}_{j_i}\mathcal{S}^{-}_{j_i}
\end{bmatrix}
\end{align}
\end{subequations}
where $\mathcal{S}^{\pm}_{j_i}$ represent the collective operators. Adding Eqs.(\ref{Eq:U}) and (\ref{Eq:U_Sq}) to Eq.(\ref{Eq:U_Approx}), the following expression is obtained.
\begin{equation}
\mathcal{U}(\tau)=
\begin{bmatrix}
\mathbb{1}-\frac{\tau^2}{2}\mathcal{S}^{-}_{j_i}\mathcal{S}^{+}_{j_i} & -i\tau \mathcal{S}^{-}_{j_i} \\
-i\tau\mathcal{S}^{+}_{j_i} & \mathbb{1}-\frac{\tau^2}{2}\mathcal{S}^{+}_{j_i}\mathcal{S}^{-}_{j_i}
\end{bmatrix},
\end{equation}
It is the time-evolution operator in matrix form as given in Eq.(\ref{Eq:Unitary Expansion}). 

\section{The Master Equation}\label{AppB}
Rewriting Eq.(\ref{Eq:U_Approx}) as $\mathcal{U}(\tau)\simeq \mathbb{1}-U_1(\tau)-U_2(\tau)$ with $U_1(\tau)=i\tau U$ and $U_2(\tau)=\tau^2 U^2/2$, and inserting into Eq.(\ref{Eq:MicroMaser}), one obtains 
\begin{align}
\dot{\varrho}_0(t)=& r\text{Tr}_{\mathcal{R}_i}[U_1(\tau)\varrho(t)U_1^{\dagger}(\tau)-U_1(\tau)\varrho(t)\nonumber\\
&-U_2(\tau)\varrho(t)-\varrho(t) U_1^{\dagger}(\tau)-\varrho(t) U_2^{\dagger}(\tau)]. 
\end{align}
Here, we have eliminated the terms higher than second order in $\tau$ such as $U_2(\tau)\varrho(t) U_2^{\dagger}(\tau)\propto \tau^4$ or $U_1(\tau)\varrho(t) U_2^{\dagger}(\tau)\propto \tau^3$. We obtain the explicit version of the master equation for the probe qubit as: by tracing out the degrees of freedom in the information environment and exploiting the linearity and cyclic features of the trace operation.   
\begin{widetext}
\begin{align}
\dot{\varrho}_0=&-ir\tau\left[\sum_i^N J_i\left(\langle\sigma_i^{-}\rangle\sigma_0^{+}+\langle\sigma_i^{+}\rangle\sigma_0^{-}\right),\varrho\right]+\frac{r\tau^2}{2}\sum_{i=1}^N J_i^2\left(\langle \sigma_i^{+}\sigma_i^{-}\rangle\mathcal{L}[\sigma_0^{+}]+\langle \sigma_i^{-}\sigma_i^{+}\rangle\mathcal{L}[\sigma_0^{-}]\right)\nonumber\\
&+2r\tau^2\sum_{i<j}^{N'} J_i J_j \left(\langle \sigma_i^{+}\rangle\langle\sigma_j^{+}\rangle \mathcal{L}_s[\sigma_0^{-}]+\langle \sigma_i^{-}\rangle\langle\sigma_j^{-}\rangle \mathcal{L}_s[\sigma_0^{+}]\right).
\end{align}
\end{widetext} 

Considering the general form of the probe qubit $\varrho_0(t)=p_e(t)\ket{e}\bra{e}+p_g(t)\ket{g}\bra{g}+(c(t)\ket{e}\bra{g}+\text{H.c.})$ in the standard basis, Bloch equations read
\begin{align}\label{Eq:Bloch}
\langle{\dot{\sigma_{x}}}\rangle=&-\left( \gamma^+_3+\gamma^-_4 \right)\langle\sigma_{x}\rangle+i\left( \gamma^-_1-\gamma^+_2 \right)\langle\sigma_{z}\rangle\nonumber\\
&+\gamma^-_5 c^*+\gamma^+_6 c\nonumber\\
\langle{\dot{\sigma_{y}}}\rangle=&-\left( \gamma^+_3+\gamma^-_4 \right)\langle\sigma_{y}\rangle-\left( \gamma^-_1+\gamma^+_2 \right)\langle\sigma_{z}\rangle\nonumber\\
&+i\gamma^-_5 c^*-i\gamma^+_6 c\nonumber\\
\langle\dot{\sigma_{z}}\rangle=&-2((\gamma^+_3-\gamma^-_4)\langle\sigma_{z}\rangle+\gamma^+_3-\gamma^-_4\nonumber\\
&+i\gamma^-_1c^*+i\gamma^+_2c)
\end{align}
where, respectively, $\gamma^-_1=r\tau\sum_{i=1}^N J_i\langle\sigma_{i}^-\rangle$, $\gamma^+_2=r\tau\sum_{i=1}^N J_i\langle\sigma_{i}^+\rangle$, $\gamma^+_3=\frac{r\tau^2}{2}\sum_{i=1}^N J_i^2\langle\sigma_{i}^+\sigma_{i}^-\rangle$, $\gamma^-_4=\frac{r\tau^2}{2}\sum_{i=1}^N J_i^2\langle\sigma_{i}^-\sigma_{i}^+\rangle$, $\gamma^-_5=2r\tau^2\sum_{i<l}^{N'}J_i J_j \langle\sigma_{i}^-\rangle\langle\sigma_{j}^-\rangle$ and $\gamma^+_6=2r\tau^2\sum_{i<j}^{N'}J_i J_j \langle\sigma_{i}^+\rangle\langle\sigma_{j}^+\rangle$  . Solutions of the Bloch equations in the steady state found to be 

\begin{align}\label{Eq:Bloch_ss}
&\langle\sigma_{x}\rangle^{ss}=\frac{i(\gamma_1^- -\gamma_2^+)}{J_{\sum}}\sum_i^N J_i^2\langle\sigma_z\rangle_i\nonumber\\
&\langle\sigma_{y}\rangle^{ss}=\frac{-(\gamma_1^- +\gamma_2^+)}{J_{\sum}}\sum_i^N J_i^2\langle\sigma_z\rangle_i\nonumber\\
&\langle\sigma_{z}\rangle^{ss}=\frac{1}{J_{\sum}}\sum_i^N J_i^2\langle\sigma_z\rangle_i\nonumber\\
\end{align}

\section{Fisher Information}\label{AppC}

$\varrho_0^{\text{ss}}$ is calculated using the variables $\theta$ and $\phi$ and is denoted as follows when the values in Eq.~(\ref{Eq:RhoR}) are substituted in Eq.~(\ref{Eq:Steady}):
\begin{widetext}
\begin{align}\label{Eq:STTP}
\varrho_0^{\text{ss}}(\theta_{i}, \phi_{i})&=
\begin{bmatrix}
\frac{1}{2J_{\sum}}  \sum_{i=1}^N \left( J_i^2+J_i^2\cos \theta_{i} \right) & \frac{i \tau r}{2J_{\sum}} \sum_{i,j=1}^N J_i J_j^2 \sin \theta_{i} \cos \theta_{j} e^{-i\phi_{i}} \\
-\frac{i \tau r}{2J_{\sum}} \sum_{i,j=1}^N J_i J_j^2 \sin \theta_{i} \cos \theta_{j} e^{i\phi_{i}}& \frac{1}{2J_{\sum}}  \sum_{i=1}^N \left( J_i^2 - J_i^2\cos \theta_{i} \right)
\end{bmatrix}
=
\begin{bmatrix}
[\varrho_0^{\text{ss}}]_{11} & C_{\theta,\phi} \\
C_{\theta,\phi}^* & [\varrho_0^{\text{ss}}]_{22}
\end{bmatrix}
\end{align}
\end{widetext} 
where $ C_{\theta,\phi}^*$ is the complex conjugate of $ C_{\theta,\phi}$.  Let $\lambda=\phi$ according to Eq.~(\ref{Eq:Fisher}). Therefore, if we take the partial derivative of Eq.~(\ref{Eq:STTP}) with respect to $\phi$, we get as follows
\begin{align}\label{Eq:DPHI}
\frac{\partial}{\partial \phi}\varrho_0^{\text{ss}}(\theta_{i}, \phi_{i})&=
\begin{bmatrix}
0 & \partial_{\phi} C_{\theta,\phi} \\
\partial_{\phi} C_{\theta,\phi}^* & 0
\end{bmatrix}
\end{align}
where $ \partial_{\phi} C_{\theta,\phi}=iC_{\theta,\phi} $ and $ \partial_{\phi} C_{\theta,\phi}^*=-iC_{\theta,\phi}^*$. 
\begin{subequations}
\begin{align}\label{Eq:Fisher1}
\text{Tr}[(\partial_{\phi}\varrho_0^{\text{ss}})^2]=2|C_{\theta,\phi}|^2
\end{align}
\begin{align}\label{Eq:Fisher2}
\text{det}\varrho_0^{\text{ss}}=[\varrho_0^{\text{ss}}]_{11}[\varrho_0^{\text{ss}}]_{22}-|C_{\theta,\phi}|^2
\end{align}
\begin{align}\label{Eq:Fisher3}
\text{Tr}[(\varrho_0^{\text{ss}} \partial_{\phi}\varrho_0^{\text{ss}})^2]=2|C_{\theta,\phi}|^2([\varrho_0^{\text{ss}}]_{11}[\varrho_0^{\text{ss}}]_{22}-|C_{\theta,\phi}|^2)
\end{align}
\end{subequations}

When we write Eqs.~(\ref{Eq:Fisher1})-(\ref{Eq:Fisher3}) in  Eq.~(\ref{Eq:Fisher}), we get the following expression as
\begin{align}
\mathcal{F}_{\phi}=4|C_{\theta,\phi}|^2.
\end{align}
Here $|C_{\theta,\phi}|^2=(C_{\theta,\phi}^*)(C_{\theta,\phi})$. Finally, QFI related to the $\phi$ is as 
\begin{align}\label{Eq:FisherPhi}
\mathcal{F}_{\phi}=\left(\frac{ \tau r}{J_{\sum}}\right)^2\sum_{\substack{i,j,\\k,l=1}}^N J_i J_j^2J_k J_l^2 \sin \theta_{i} \cos \theta_{j} \sin \theta_{k} \cos \theta_{l} e^{i(\phi_{i}-\phi_{k})}.
\end{align}

Let $\lambda=\theta$ according to Eq.~(\ref{Eq:Fisher}). Therefore, if we take the partial derivative of Eq.~(\ref{Eq:STTP}) with respect to $\theta$, one obtains
\begin{widetext}
\begin{align}\label{Eq:DTHETA}
\frac{\partial}{\partial \theta}\varrho_0^{\text{ss}}(\theta_{i}, \phi_{i})&=
\begin{bmatrix}
-\frac{1}{2J_{\sum}}  \sum_{i=1}^N  J_i^2\sin \theta_{i} & \frac{i \tau r}{2J_{\sum}} \sum_{i,j=1}^N J_i J_j^2 \cos(\theta_{i}+ \theta_{j}) e^{-i\phi_{i}}  \\
 -\frac{i \tau r}{2J_{\sum}} \sum_{i,j=1}^N J_i J_j^2 \cos(\theta_{i}+ \theta_{j}) e^{i\phi_{i}}& \frac{1}{2J_{\sum}}  \sum_{i=1}^N  J_i^2\sin \theta_{i} 
\end{bmatrix}
=
\begin{bmatrix}
[\partial_{\theta} \varrho_0^{\text{ss}}]_{11} & \partial_{\theta} C_{\theta,\phi} \\
\partial_{\theta} C_{\theta,\phi}^* & [\partial_{\theta}  \varrho_0^{\text{ss}}]_{22}
\end{bmatrix}.
\end{align}
\end{widetext} 
By following the steps we used for $\phi$  above,  also quantum Fisher information can be obtained according to $\theta$. QFI expressions based on amplitude parameters for a single rezervuar of information are given in Eqs.~(\ref{Eq:QFIT}) and~(\ref{Eq:QFIP}).

\section{Derivation of the cost function}\label{AppD} 
Let's edit Eq.~(\ref{Eq:delta_nu}) for $J$

\begin{equation}\label{Eq:deltag}
\delta {J}_i=-\eta\frac{\partial C}{\partial {J}_i}.
\end{equation}

When we consider the partial derivative of the cost function with respect to the coupling constant $J$, we arrive at Eq.~(\ref{Eq:deltac}).
\begin{equation}\label{Eq:deltac}
\frac{\partial C}{\partial {J}_i}=(\langle \sigma_z^0\rangle_{des}^{ss}-\langle \sigma_z^0\rangle_{act}^{ss})(-\frac{\partial \langle \sigma_z^0\rangle_{act}^{ss}}{\partial {J}_i})
\end{equation}

In the scenario we are using right now, there are two information reservoirs that correspond to specific magnetizations. The actual steady state magnetization (Eq.~(\ref{Eq:Bloch_ss})) thus has the value

\begin{equation}\label{Eq:Actual}
A=\langle \sigma_z^0\rangle_{act}^{ss}=\frac{J_{1}^{2}\langle \sigma_z^1\rangle+J_{2}^{2}\langle \sigma_z^2\rangle}{J_{1}^{2}+J_{2}^{2}}.
\end{equation}
The partial derivatives with regard to $J_1$ and $J_2$ separately were calculated using the formula to determine the cost function, and they were as follows:
\begin{align}\label{Eq:pdg1_g2}
&\frac{\partial A}{\partial {J}_1}=\frac{2J_{1}\langle \sigma_z^1\rangle (J_{1}^{2}+J_{2}^{2})-2J_{1}(J_{1}^{2}\langle \sigma_z^1\rangle+J_{2}^{2}\langle \sigma_z^2\rangle)}{(J_{1}^{2}+J_{2}^{2})^{2}}\nonumber\\
&\frac{\partial A}{\partial {J}_2}=\frac{2J_{2}\langle \sigma_z^2\rangle (J_{1}^{2}+J_{2}^{2})-2J_{2}(J_{1}^{2}\langle \sigma_z^1\rangle+J_{2}^{2}\langle \sigma_z^2\rangle)}{(J_{1}^{2}+J_{2}^{2})^{2}}
\end{align}
In our case, the cost function's constant value for the desired magnetization is $\langle \sigma_z^0\rangle_{des}^{ss}=0.4$. After substituting Eqs.~(\ref{Eq:Actual}) and~(\ref{Eq:pdg1_g2}) in Eq.~(\ref{Eq:deltac}), the expression obtained after substituting them in Eq.~(\ref{Eq:deltag}), Eq.~(\ref{Eq:training}) becomes as follows:
\begin{align}\label{Eq:training_1}
&\bf{(J_1)}_{k+1}=\bf{(J_1)}_{k}+\delta\bf{(J_1)}_{k}\nonumber\\
&\bf{(J_2)}_{k+1}=\bf{(J_2)}_{k}+\delta\bf{(J_2)}_{k}.
\end{align}

\balance

\begin{thebibliography}{60}%
\makeatletter
\providecommand \@ifxundefined [1]{%
 \@ifx{#1\undefined}
}%
\providecommand \@ifnum [1]{%
 \ifnum #1\expandafter \@firstoftwo
 \else \expandafter \@secondoftwo
 \fi
}%
\providecommand \@ifx [1]{%
 \ifx #1\expandafter \@firstoftwo
 \else \expandafter \@secondoftwo
 \fi
}%
\providecommand \natexlab [1]{#1}%
\providecommand \enquote  [1]{``#1''}%
\providecommand \bibnamefont  [1]{#1}%
\providecommand \bibfnamefont [1]{#1}%
\providecommand \citenamefont [1]{#1}%
\providecommand \href@noop [0]{\@secondoftwo}%
\providecommand \href [0]{\begingroup \@sanitize@url \@href}%
\providecommand \@href[1]{\@@startlink{#1}\@@href}%
\providecommand \@@href[1]{\endgroup#1\@@endlink}%
\providecommand \@sanitize@url [0]{\catcode `\\12\catcode `\$12\catcode
  `\&12\catcode `\#12\catcode `\^12\catcode `\_12\catcode `\%12\relax}%
\providecommand \@@startlink[1]{}%
\providecommand \@@endlink[0]{}%
\providecommand \url  [0]{\begingroup\@sanitize@url \@url }%
\providecommand \@url [1]{\endgroup\@href {#1}{\urlprefix }}%
\providecommand \urlprefix  [0]{URL }%
\providecommand \Eprint [0]{\href }%
\providecommand \doibase [0]{https://doi.org/}%
\providecommand \selectlanguage [0]{\@gobble}%
\providecommand \bibinfo  [0]{\@secondoftwo}%
\providecommand \bibfield  [0]{\@secondoftwo}%
\providecommand \translation [1]{[#1]}%
\providecommand \BibitemOpen [0]{}%
\providecommand \bibitemStop [0]{}%
\providecommand \bibitemNoStop [0]{.\EOS\space}%
\providecommand \EOS [0]{\spacefactor3000\relax}%
\providecommand \BibitemShut  [1]{\csname bibitem#1\endcsname}%
\let\auto@bib@innerbib\@empty
\bibitem [{\citenamefont {Biamonte}\ \emph {et~al.}(2017)\citenamefont
  {Biamonte}, \citenamefont {Wittek}, \citenamefont {Pancotti}, \citenamefont
  {Rebentrost}, \citenamefont {Wiebe},\ and\ \citenamefont
  {Lloyd}}]{biamonte_quantum_2017}%
  \BibitemOpen
  \bibfield  {author} {\bibinfo {author} {\bibfnamefont {J.}~\bibnamefont
  {Biamonte}}, \bibinfo {author} {\bibfnamefont {P.}~\bibnamefont {Wittek}},
  \bibinfo {author} {\bibfnamefont {N.}~\bibnamefont {Pancotti}}, \bibinfo
  {author} {\bibfnamefont {P.}~\bibnamefont {Rebentrost}}, \bibinfo {author}
  {\bibfnamefont {N.}~\bibnamefont {Wiebe}},\ and\ \bibinfo {author}
  {\bibfnamefont {S.}~\bibnamefont {Lloyd}},\ }\href
  {https://doi.org/10.1038/nature23474} {\bibfield  {journal} {\bibinfo
  {journal} {Nature}\ }\textbf {\bibinfo {volume} {549}},\ \bibinfo {pages}
  {195} (\bibinfo {year} {2017})}\BibitemShut {NoStop}%
\bibitem [{\citenamefont {Aïmeur}\ \emph {et~al.}(2013)\citenamefont
  {Aïmeur}, \citenamefont {Brassard},\ and\ \citenamefont
  {Gambs}}]{aimeur_quantum_2013}%
  \BibitemOpen
  \bibfield  {author} {\bibinfo {author} {\bibfnamefont {E.}~\bibnamefont
  {Aïmeur}}, \bibinfo {author} {\bibfnamefont {G.}~\bibnamefont {Brassard}},\
  and\ \bibinfo {author} {\bibfnamefont {S.}~\bibnamefont {Gambs}},\ }\href
  {https://doi.org/10.1007/s10994-012-5316-5} {\bibfield  {journal} {\bibinfo
  {journal} {Machine Learning}\ }\textbf {\bibinfo {volume} {90}},\ \bibinfo
  {pages} {261} (\bibinfo {year} {2013})}\BibitemShut {NoStop}%
\bibitem [{\citenamefont {Paparo}\ \emph {et~al.}(2014)\citenamefont {Paparo},
  \citenamefont {Dunjko}, \citenamefont {Makmal}, \citenamefont
  {Martin-Delgado},\ and\ \citenamefont {Briegel}}]{paparo_quantum_2014}%
  \BibitemOpen
  \bibfield  {author} {\bibinfo {author} {\bibfnamefont {G.~D.}\ \bibnamefont
  {Paparo}}, \bibinfo {author} {\bibfnamefont {V.}~\bibnamefont {Dunjko}},
  \bibinfo {author} {\bibfnamefont {A.}~\bibnamefont {Makmal}}, \bibinfo
  {author} {\bibfnamefont {M.~A.}\ \bibnamefont {Martin-Delgado}},\ and\
  \bibinfo {author} {\bibfnamefont {H.~J.}\ \bibnamefont {Briegel}},\ }\href
  {https://doi.org/10.1103/PhysRevX.4.031002} {\bibfield  {journal} {\bibinfo
  {journal} {Physical Review X}\ }\textbf {\bibinfo {volume} {4}},\ \bibinfo
  {pages} {031002} (\bibinfo {year} {2014})}\BibitemShut {NoStop}%
\bibitem [{\citenamefont {Dunjko}\ \emph {et~al.}(2016)\citenamefont {Dunjko},
  \citenamefont {Taylor},\ and\ \citenamefont
  {Briegel}}]{dunjko_quantum-enhanced_2016}%
  \BibitemOpen
  \bibfield  {author} {\bibinfo {author} {\bibfnamefont {V.}~\bibnamefont
  {Dunjko}}, \bibinfo {author} {\bibfnamefont {J.~M.}\ \bibnamefont {Taylor}},\
  and\ \bibinfo {author} {\bibfnamefont {H.~J.}\ \bibnamefont {Briegel}},\
  }\href {https://doi.org/10.1103/PhysRevLett.117.130501} {\bibfield  {journal}
  {\bibinfo  {journal} {Physical Review Letters}\ }\textbf {\bibinfo {volume}
  {117}},\ \bibinfo {pages} {130501} (\bibinfo {year} {2016})}\BibitemShut
  {NoStop}%
\bibitem [{\citenamefont {Alvarez-Rodriguez}\ \emph {et~al.}(2017)\citenamefont
  {Alvarez-Rodriguez}, \citenamefont {Lamata}, \citenamefont
  {Escandell-Montero}, \citenamefont {Martín-Guerrero},\ and\ \citenamefont
  {Solano}}]{alvarez-rodriguez_supervised_2017}%
  \BibitemOpen
  \bibfield  {author} {\bibinfo {author} {\bibfnamefont {U.}~\bibnamefont
  {Alvarez-Rodriguez}}, \bibinfo {author} {\bibfnamefont {L.}~\bibnamefont
  {Lamata}}, \bibinfo {author} {\bibfnamefont {P.}~\bibnamefont
  {Escandell-Montero}}, \bibinfo {author} {\bibfnamefont {J.~D.}\ \bibnamefont
  {Martín-Guerrero}},\ and\ \bibinfo {author} {\bibfnamefont {E.}~\bibnamefont
  {Solano}},\ }\href {https://doi.org/10.1038/s41598-017-13378-0} {\bibfield
  {journal} {\bibinfo  {journal} {Scientific Reports}\ }\textbf {\bibinfo
  {volume} {7}},\ \bibinfo {pages} {13645} (\bibinfo {year}
  {2017})}\BibitemShut {NoStop}%
\bibitem [{\citenamefont {Mitarai}\ \emph {et~al.}(2018)\citenamefont
  {Mitarai}, \citenamefont {Negoro}, \citenamefont {Kitagawa},\ and\
  \citenamefont {Fujii}}]{mitarai_quantum_2018}%
  \BibitemOpen
  \bibfield  {author} {\bibinfo {author} {\bibfnamefont {K.}~\bibnamefont
  {Mitarai}}, \bibinfo {author} {\bibfnamefont {M.}~\bibnamefont {Negoro}},
  \bibinfo {author} {\bibfnamefont {M.}~\bibnamefont {Kitagawa}},\ and\
  \bibinfo {author} {\bibfnamefont {K.}~\bibnamefont {Fujii}},\ }\href
  {https://doi.org/10.1103/PhysRevA.98.032309} {\bibfield  {journal} {\bibinfo
  {journal} {Physical Review A}\ }\textbf {\bibinfo {volume} {98}},\ \bibinfo
  {pages} {032309} (\bibinfo {year} {2018})}\BibitemShut {NoStop}%
\bibitem [{\citenamefont {Schuld}\ \emph {et~al.}(2015)\citenamefont {Schuld},
  \citenamefont {Sinayskiy},\ and\ \citenamefont
  {Petruccione}}]{schuld_simulating_2015}%
  \BibitemOpen
  \bibfield  {author} {\bibinfo {author} {\bibfnamefont {M.}~\bibnamefont
  {Schuld}}, \bibinfo {author} {\bibfnamefont {I.}~\bibnamefont {Sinayskiy}},\
  and\ \bibinfo {author} {\bibfnamefont {F.}~\bibnamefont {Petruccione}},\
  }\href {https://doi.org/10.1016/j.physleta.2014.11.061} {\bibfield  {journal}
  {\bibinfo  {journal} {Physics Letters A}\ }\textbf {\bibinfo {volume}
  {379}},\ \bibinfo {pages} {660} (\bibinfo {year} {2015})}\BibitemShut
  {NoStop}%
\bibitem [{\citenamefont {Wan}\ \emph {et~al.}(2017)\citenamefont {Wan},
  \citenamefont {Dahlsten}, \citenamefont {Kristjánsson}, \citenamefont
  {Gardner},\ and\ \citenamefont {Kim}}]{wan_quantum_2017}%
  \BibitemOpen
  \bibfield  {author} {\bibinfo {author} {\bibfnamefont {K.~H.}\ \bibnamefont
  {Wan}}, \bibinfo {author} {\bibfnamefont {O.}~\bibnamefont {Dahlsten}},
  \bibinfo {author} {\bibfnamefont {H.}~\bibnamefont {Kristjánsson}}, \bibinfo
  {author} {\bibfnamefont {R.}~\bibnamefont {Gardner}},\ and\ \bibinfo {author}
  {\bibfnamefont {M.~S.}\ \bibnamefont {Kim}},\ }\href
  {https://doi.org/10.1038/s41534-017-0032-4} {\bibfield  {journal} {\bibinfo
  {journal} {npj Quantum Information}\ }\textbf {\bibinfo {volume} {3}},\
  \bibinfo {pages} {1} (\bibinfo {year} {2017})}\BibitemShut {NoStop}%
\bibitem [{\citenamefont {Schuld}\ and\ \citenamefont
  {Petruccione}(2018)}]{schuld_quantum_2018}%
  \BibitemOpen
  \bibfield  {author} {\bibinfo {author} {\bibfnamefont {M.}~\bibnamefont
  {Schuld}}\ and\ \bibinfo {author} {\bibfnamefont {F.}~\bibnamefont
  {Petruccione}},\ }\href {https://doi.org/10.1038/s41598-018-20403-3}
  {\bibfield  {journal} {\bibinfo  {journal} {Scientific Reports}\ }\textbf
  {\bibinfo {volume} {8}},\ \bibinfo {pages} {2772} (\bibinfo {year}
  {2018})}\BibitemShut {NoStop}%
\bibitem [{\citenamefont {Tacchino}\ \emph {et~al.}(2019)\citenamefont
  {Tacchino}, \citenamefont {Macchiavello}, \citenamefont {Gerace},\ and\
  \citenamefont {Bajoni}}]{tacchino_artificial_2019}%
  \BibitemOpen
  \bibfield  {author} {\bibinfo {author} {\bibfnamefont {F.}~\bibnamefont
  {Tacchino}}, \bibinfo {author} {\bibfnamefont {C.}~\bibnamefont
  {Macchiavello}}, \bibinfo {author} {\bibfnamefont {D.}~\bibnamefont
  {Gerace}},\ and\ \bibinfo {author} {\bibfnamefont {D.}~\bibnamefont
  {Bajoni}},\ }\href {https://doi.org/10.1038/s41534-019-0140-4} {\bibfield
  {journal} {\bibinfo  {journal} {npj Quantum Information}\ }\textbf {\bibinfo
  {volume} {5}},\ \bibinfo {pages} {1} (\bibinfo {year} {2019})}\BibitemShut
  {NoStop}%
\bibitem [{\citenamefont {Torrontegui}\ and\ \citenamefont
  {García-Ripoll}(2019)}]{torrontegui_unitary_2019}%
  \BibitemOpen
  \bibfield  {author} {\bibinfo {author} {\bibfnamefont {E.}~\bibnamefont
  {Torrontegui}}\ and\ \bibinfo {author} {\bibfnamefont {J.~J.}\ \bibnamefont
  {García-Ripoll}},\ }\href {https://doi.org/10.1209/0295-5075/125/30004}
  {\bibfield  {journal} {\bibinfo  {journal} {EPL (Europhysics Letters)}\
  }\textbf {\bibinfo {volume} {125}},\ \bibinfo {pages} {30004} (\bibinfo
  {year} {2019})}\BibitemShut {NoStop}%
\bibitem [{\citenamefont {Beer}\ \emph {et~al.}(2020)\citenamefont {Beer},
  \citenamefont {Bondarenko}, \citenamefont {Farrelly}, \citenamefont
  {Osborne}, \citenamefont {Salzmann}, \citenamefont {Scheiermann},\ and\
  \citenamefont {Wolf}}]{beer_training_2020}%
  \BibitemOpen
  \bibfield  {author} {\bibinfo {author} {\bibfnamefont {K.}~\bibnamefont
  {Beer}}, \bibinfo {author} {\bibfnamefont {D.}~\bibnamefont {Bondarenko}},
  \bibinfo {author} {\bibfnamefont {T.}~\bibnamefont {Farrelly}}, \bibinfo
  {author} {\bibfnamefont {T.~J.}\ \bibnamefont {Osborne}}, \bibinfo {author}
  {\bibfnamefont {R.}~\bibnamefont {Salzmann}}, \bibinfo {author}
  {\bibfnamefont {D.}~\bibnamefont {Scheiermann}},\ and\ \bibinfo {author}
  {\bibfnamefont {R.}~\bibnamefont {Wolf}},\ }\href
  {https://doi.org/10.1038/s41467-020-14454-2} {\bibfield  {journal} {\bibinfo
  {journal} {Nature Communications}\ }\textbf {\bibinfo {volume} {11}},\
  \bibinfo {pages} {808} (\bibinfo {year} {2020})}\BibitemShut {NoStop}%
\bibitem [{\citenamefont {Preskill}(2018)}]{preskill_quantum_2018}%
  \BibitemOpen
  \bibfield  {author} {\bibinfo {author} {\bibfnamefont {J.}~\bibnamefont
  {Preskill}},\ }\href {https://quantum-journal.org/papers/q-2018-08-06-79/}
  {\bibfield  {journal} {\bibinfo  {journal} {Quantum}\ }\textbf {\bibinfo
  {volume} {2}},\ \bibinfo {pages} {79} (\bibinfo {year} {2018})}\BibitemShut
  {NoStop}%
\bibitem [{\citenamefont {Verstraete}\ \emph {et~al.}(2009)\citenamefont
  {Verstraete}, \citenamefont {Wolf},\ and\ \citenamefont
  {Ignacio~Cirac}}]{verstraete_quantum_2009}%
  \BibitemOpen
  \bibfield  {author} {\bibinfo {author} {\bibfnamefont {F.}~\bibnamefont
  {Verstraete}}, \bibinfo {author} {\bibfnamefont {M.~M.}\ \bibnamefont
  {Wolf}},\ and\ \bibinfo {author} {\bibfnamefont {J.}~\bibnamefont
  {Ignacio~Cirac}},\ }\href {https://doi.org/10.1038/nphys1342} {\bibfield
  {journal} {\bibinfo  {journal} {Nature Physics}\ }\textbf {\bibinfo {volume}
  {5}},\ \bibinfo {pages} {633} (\bibinfo {year} {2009})}\BibitemShut {NoStop}%
\bibitem [{\citenamefont {Sinayskiy}\ and\ \citenamefont
  {Petruccione}(2012)}]{sinayskiy_efficiency_2012}%
  \BibitemOpen
  \bibfield  {author} {\bibinfo {author} {\bibfnamefont {I.}~\bibnamefont
  {Sinayskiy}}\ and\ \bibinfo {author} {\bibfnamefont {F.}~\bibnamefont
  {Petruccione}},\ }\href {https://doi.org/10.1007/s11128-012-0426-3}
  {\bibfield  {journal} {\bibinfo  {journal} {Quantum Information Processing}\
  }\textbf {\bibinfo {volume} {11}},\ \bibinfo {pages} {1301} (\bibinfo {year}
  {2012})}\BibitemShut {NoStop}%
\bibitem [{\citenamefont {Schuld}\ \emph {et~al.}(2014)\citenamefont {Schuld},
  \citenamefont {Sinayskiy},\ and\ \citenamefont
  {Petruccione}}]{schuld_quantum_2014}%
  \BibitemOpen
  \bibfield  {author} {\bibinfo {author} {\bibfnamefont {M.}~\bibnamefont
  {Schuld}}, \bibinfo {author} {\bibfnamefont {I.}~\bibnamefont {Sinayskiy}},\
  and\ \bibinfo {author} {\bibfnamefont {F.}~\bibnamefont {Petruccione}},\
  }\href {https://doi.org/10.1103/PhysRevA.89.032333} {\bibfield  {journal}
  {\bibinfo  {journal} {Physical Review A}\ }\textbf {\bibinfo {volume} {89}},\
  \bibinfo {pages} {032333} (\bibinfo {year} {2014})}\BibitemShut {NoStop}%
\bibitem [{\citenamefont {Song}\ \emph {et~al.}(2015)\citenamefont {Song},
  \citenamefont {Di}, \citenamefont {Xia}, \citenamefont {Sun},\ and\
  \citenamefont {Jiang}}]{song_implementation_2015}%
  \BibitemOpen
  \bibfield  {author} {\bibinfo {author} {\bibfnamefont {J.}~\bibnamefont
  {Song}}, \bibinfo {author} {\bibfnamefont {J.-Y.}\ \bibnamefont {Di}},
  \bibinfo {author} {\bibfnamefont {Y.}~\bibnamefont {Xia}}, \bibinfo {author}
  {\bibfnamefont {X.-D.}\ \bibnamefont {Sun}},\ and\ \bibinfo {author}
  {\bibfnamefont {Y.-Y.}\ \bibnamefont {Jiang}},\ }\href
  {https://doi.org/10.1038/srep10656} {\bibfield  {journal} {\bibinfo
  {journal} {Scientific Reports}\ }\textbf {\bibinfo {volume} {5}},\ \bibinfo
  {pages} {10656} (\bibinfo {year} {2015})}\BibitemShut {NoStop}%
\bibitem [{\citenamefont {Marshall}\ \emph {et~al.}(2019)\citenamefont
  {Marshall}, \citenamefont {Campos~Venuti},\ and\ \citenamefont
  {Zanardi}}]{marshall_classifying_2019}%
  \BibitemOpen
  \bibfield  {author} {\bibinfo {author} {\bibfnamefont {J.}~\bibnamefont
  {Marshall}}, \bibinfo {author} {\bibfnamefont {L.}~\bibnamefont
  {Campos~Venuti}},\ and\ \bibinfo {author} {\bibfnamefont {P.}~\bibnamefont
  {Zanardi}},\ }\href {https://doi.org/10.1103/PhysRevA.99.032330} {\bibfield
  {journal} {\bibinfo  {journal} {Physical Review A}\ }\textbf {\bibinfo
  {volume} {99}},\ \bibinfo {pages} {032330} (\bibinfo {year}
  {2019})}\BibitemShut {NoStop}%
\bibitem [{\citenamefont {Blume-Kohout}\ and\ \citenamefont
  {Zurek}(2005)}]{blume-kohout_simple_2005}%
  \BibitemOpen
  \bibfield  {author} {\bibinfo {author} {\bibfnamefont {R.}~\bibnamefont
  {Blume-Kohout}}\ and\ \bibinfo {author} {\bibfnamefont {W.~H.}\ \bibnamefont
  {Zurek}},\ }\href {https://doi.org/10.1007/s10701-005-7352-5} {\bibfield
  {journal} {\bibinfo  {journal} {Foundations of Physics}\ }\textbf {\bibinfo
  {volume} {35}},\ \bibinfo {pages} {1857} (\bibinfo {year}
  {2005})}\BibitemShut {NoStop}%
\bibitem [{\citenamefont {Zwolak}\ and\ \citenamefont
  {Zurek}(2017)}]{zwolak_redundancy_2017}%
  \BibitemOpen
  \bibfield  {author} {\bibinfo {author} {\bibfnamefont {M.}~\bibnamefont
  {Zwolak}}\ and\ \bibinfo {author} {\bibfnamefont {W.~H.}\ \bibnamefont
  {Zurek}},\ }\href {https://doi.org/10.1103/PhysRevA.95.030101} {\bibfield
  {journal} {\bibinfo  {journal} {Physical Review A}\ }\textbf {\bibinfo
  {volume} {95}},\ \bibinfo {pages} {030101} (\bibinfo {year}
  {2017})}\BibitemShut {NoStop}%
\bibitem [{\citenamefont {Deffner}\ and\ \citenamefont
  {Jarzynski}(2013)}]{deffner_information_2013}%
  \BibitemOpen
  \bibfield  {author} {\bibinfo {author} {\bibfnamefont {S.}~\bibnamefont
  {Deffner}}\ and\ \bibinfo {author} {\bibfnamefont {C.}~\bibnamefont
  {Jarzynski}},\ }\href {https://doi.org/10.1103/PhysRevX.3.041003} {\bibfield
  {journal} {\bibinfo  {journal} {Physical Review X}\ }\textbf {\bibinfo
  {volume} {3}},\ \bibinfo {pages} {041003} (\bibinfo {year}
  {2013})}\BibitemShut {NoStop}%
\bibitem [{\citenamefont {Deffner}(2013)}]{deffner_information-driven_2013}%
  \BibitemOpen
  \bibfield  {author} {\bibinfo {author} {\bibfnamefont {S.}~\bibnamefont
  {Deffner}},\ }\href {https://doi.org/10.1103/PhysRevE.88.062128} {\bibfield
  {journal} {\bibinfo  {journal} {Physical Review E}\ }\textbf {\bibinfo
  {volume} {88}},\ \bibinfo {pages} {062128} (\bibinfo {year}
  {2013})}\BibitemShut {NoStop}%
\bibitem [{\citenamefont {Poyatos}\ \emph {et~al.}(1996)\citenamefont
  {Poyatos}, \citenamefont {Cirac},\ and\ \citenamefont
  {Zoller}}]{poyatos_quantum_1996}%
  \BibitemOpen
  \bibfield  {author} {\bibinfo {author} {\bibfnamefont {J.~F.}\ \bibnamefont
  {Poyatos}}, \bibinfo {author} {\bibfnamefont {J.~I.}\ \bibnamefont {Cirac}},\
  and\ \bibinfo {author} {\bibfnamefont {P.}~\bibnamefont {Zoller}},\ }\href
  {https://doi.org/10.1103/PhysRevLett.77.4728} {\bibfield  {journal} {\bibinfo
   {journal} {Physical Review Letters}\ }\textbf {\bibinfo {volume} {77}},\
  \bibinfo {pages} {4728} (\bibinfo {year} {1996})}\BibitemShut {NoStop}%
\bibitem [{\citenamefont {Scarani}\ \emph {et~al.}(2002)\citenamefont
  {Scarani}, \citenamefont {Ziman}, \citenamefont {Štelmachovič},
  \citenamefont {Gisin},\ and\ \citenamefont
  {Bužek}}]{scarani_thermalizing_2002}%
  \BibitemOpen
  \bibfield  {author} {\bibinfo {author} {\bibfnamefont {V.}~\bibnamefont
  {Scarani}}, \bibinfo {author} {\bibfnamefont {M.}~\bibnamefont {Ziman}},
  \bibinfo {author} {\bibfnamefont {P.}~\bibnamefont {Štelmachovič}},
  \bibinfo {author} {\bibfnamefont {N.}~\bibnamefont {Gisin}},\ and\ \bibinfo
  {author} {\bibfnamefont {V.}~\bibnamefont {Bužek}},\ }\href
  {https://doi.org/10.1103/PhysRevLett.88.097905} {\bibfield  {journal}
  {\bibinfo  {journal} {Physical Review Letters}\ }\textbf {\bibinfo {volume}
  {88}},\ \bibinfo {pages} {097905} (\bibinfo {year} {2002})}\BibitemShut
  {NoStop}%
\bibitem [{\citenamefont {Ziman}\ \emph {et~al.}(2002)\citenamefont {Ziman},
  \citenamefont {Štelmachovič}, \citenamefont {Bužek}, \citenamefont
  {Hillery}, \citenamefont {Scarani},\ and\ \citenamefont
  {Gisin}}]{ziman_diluting_2002}%
  \BibitemOpen
  \bibfield  {author} {\bibinfo {author} {\bibfnamefont {M.}~\bibnamefont
  {Ziman}}, \bibinfo {author} {\bibfnamefont {P.}~\bibnamefont
  {Štelmachovič}}, \bibinfo {author} {\bibfnamefont {V.}~\bibnamefont
  {Bužek}}, \bibinfo {author} {\bibfnamefont {M.}~\bibnamefont {Hillery}},
  \bibinfo {author} {\bibfnamefont {V.}~\bibnamefont {Scarani}},\ and\ \bibinfo
  {author} {\bibfnamefont {N.}~\bibnamefont {Gisin}},\ }\href
  {https://doi.org/10.1103/PhysRevA.65.042105} {\bibfield  {journal} {\bibinfo
  {journal} {Physical Review A}\ }\textbf {\bibinfo {volume} {65}},\ \bibinfo
  {pages} {042105} (\bibinfo {year} {2002})}\BibitemShut {NoStop}%
\bibitem [{\citenamefont {Nagaj}\ \emph {et~al.}(2002)\citenamefont {Nagaj},
  \citenamefont {Štelmachovič}, \citenamefont {Bužek},\ and\ \citenamefont
  {Kim}}]{nagaj_quantum_2002}%
  \BibitemOpen
  \bibfield  {author} {\bibinfo {author} {\bibfnamefont {D.}~\bibnamefont
  {Nagaj}}, \bibinfo {author} {\bibfnamefont {P.}~\bibnamefont
  {Štelmachovič}}, \bibinfo {author} {\bibfnamefont {V.}~\bibnamefont
  {Bužek}},\ and\ \bibinfo {author} {\bibfnamefont {M.}~\bibnamefont {Kim}},\
  }\href {https://doi.org/10.1103/PhysRevA.66.062307} {\bibfield  {journal}
  {\bibinfo  {journal} {Physical Review A}\ }\textbf {\bibinfo {volume} {66}},\
  \bibinfo {pages} {062307} (\bibinfo {year} {2002})}\BibitemShut {NoStop}%
\bibitem [{\citenamefont {Karevski}\ and\ \citenamefont
  {Platini}(2009)}]{karevski_quantum_2009}%
  \BibitemOpen
  \bibfield  {author} {\bibinfo {author} {\bibfnamefont {D.}~\bibnamefont
  {Karevski}}\ and\ \bibinfo {author} {\bibfnamefont {T.}~\bibnamefont
  {Platini}},\ }\href {https://doi.org/10.1103/PhysRevLett.102.207207}
  {\bibfield  {journal} {\bibinfo  {journal} {Physical Review Letters}\
  }\textbf {\bibinfo {volume} {102}},\ \bibinfo {pages} {207207} (\bibinfo
  {year} {2009})}\BibitemShut {NoStop}%
\bibitem [{\citenamefont {Seah}\ \emph
  {et~al.}(2019{\natexlab{a}})\citenamefont {Seah}, \citenamefont
  {Nimmrichter},\ and\ \citenamefont {Scarani}}]{seah_nonequilibrium_2019}%
  \BibitemOpen
  \bibfield  {author} {\bibinfo {author} {\bibfnamefont {S.}~\bibnamefont
  {Seah}}, \bibinfo {author} {\bibfnamefont {S.}~\bibnamefont {Nimmrichter}},\
  and\ \bibinfo {author} {\bibfnamefont {V.}~\bibnamefont {Scarani}},\ }\href
  {https://doi.org/10.1103/PhysRevE.99.042103} {\bibfield  {journal} {\bibinfo
  {journal} {Physical Review E}\ }\textbf {\bibinfo {volume} {99}},\ \bibinfo
  {pages} {042103} (\bibinfo {year} {2019}{\natexlab{a}})}\BibitemShut
  {NoStop}%
\bibitem [{\citenamefont {Ciccarello}\ \emph {et~al.}(2013)\citenamefont
  {Ciccarello}, \citenamefont {Palma},\ and\ \citenamefont
  {Giovannetti}}]{ciccarello_collision-model-based_2013}%
  \BibitemOpen
  \bibfield  {author} {\bibinfo {author} {\bibfnamefont {F.}~\bibnamefont
  {Ciccarello}}, \bibinfo {author} {\bibfnamefont {G.~M.}\ \bibnamefont
  {Palma}},\ and\ \bibinfo {author} {\bibfnamefont {V.}~\bibnamefont
  {Giovannetti}},\ }\href {https://doi.org/10.1103/PhysRevA.87.040103}
  {\bibfield  {journal} {\bibinfo  {journal} {Physical Review A}\ }\textbf
  {\bibinfo {volume} {87}},\ \bibinfo {pages} {040103} (\bibinfo {year}
  {2013})}\BibitemShut {NoStop}%
\bibitem [{\citenamefont {Kretschmer}\ \emph {et~al.}(2016)\citenamefont
  {Kretschmer}, \citenamefont {Luoma},\ and\ \citenamefont
  {Strunz}}]{kretschmer_collision_2016}%
  \BibitemOpen
  \bibfield  {author} {\bibinfo {author} {\bibfnamefont {S.}~\bibnamefont
  {Kretschmer}}, \bibinfo {author} {\bibfnamefont {K.}~\bibnamefont {Luoma}},\
  and\ \bibinfo {author} {\bibfnamefont {W.~T.}\ \bibnamefont {Strunz}},\
  }\href {https://doi.org/10.1103/PhysRevA.94.012106} {\bibfield  {journal}
  {\bibinfo  {journal} {Physical Review A}\ }\textbf {\bibinfo {volume} {94}},\
  \bibinfo {pages} {012106} (\bibinfo {year} {2016})}\BibitemShut {NoStop}%
\bibitem [{\citenamefont {McCloskey}\ and\ \citenamefont
  {Paternostro}(2014)}]{mccloskey_non-markovianity_2014}%
  \BibitemOpen
  \bibfield  {author} {\bibinfo {author} {\bibfnamefont {R.}~\bibnamefont
  {McCloskey}}\ and\ \bibinfo {author} {\bibfnamefont {M.}~\bibnamefont
  {Paternostro}},\ }\href {https://doi.org/10.1103/PhysRevA.89.052120}
  {\bibfield  {journal} {\bibinfo  {journal} {Physical Review A}\ }\textbf
  {\bibinfo {volume} {89}},\ \bibinfo {pages} {052120} (\bibinfo {year}
  {2014})}\BibitemShut {NoStop}%
\bibitem [{\citenamefont {Türkpençe}\ \emph {et~al.}(2017)\citenamefont
  {Türkpençe}, \citenamefont {Altintas}, \citenamefont {Paternostro},\ and\
  \citenamefont {M\"{u}stecapl{\i}o\u{g}lu}}]{turkpence_photonic_2017}%
  \BibitemOpen
  \bibfield  {author} {\bibinfo {author} {\bibfnamefont {D.}~\bibnamefont
  {Türkpençe}}, \bibinfo {author} {\bibfnamefont {F.}~\bibnamefont
  {Altintas}}, \bibinfo {author} {\bibfnamefont {M.}~\bibnamefont
  {Paternostro}},\ and\ \bibinfo {author} {\bibfnamefont {\"{O}.~E.}\ \bibnamefont
  {M\"{u}stecapl{\i}o\u{g}lu}},\ }\href {https://doi.org/10.1209/0295-5075/117/50002}
  {\bibfield  {journal} {\bibinfo  {journal} {EPL (Europhysics Letters)}\
  }\textbf {\bibinfo {volume} {117}},\ \bibinfo {pages} {50002} (\bibinfo
  {year} {2017})}\BibitemShut {NoStop}%
\bibitem [{\citenamefont {Seah}\ \emph
  {et~al.}(2019{\natexlab{b}})\citenamefont {Seah}, \citenamefont
  {Nimmrichter}, \citenamefont {Grimmer}, \citenamefont {Santos}, \citenamefont
  {Scarani},\ and\ \citenamefont {Landi}}]{seah_collisional_2019}%
  \BibitemOpen
  \bibfield  {author} {\bibinfo {author} {\bibfnamefont {S.}~\bibnamefont
  {Seah}}, \bibinfo {author} {\bibfnamefont {S.}~\bibnamefont {Nimmrichter}},
  \bibinfo {author} {\bibfnamefont {D.}~\bibnamefont {Grimmer}}, \bibinfo
  {author} {\bibfnamefont {J.~P.}\ \bibnamefont {Santos}}, \bibinfo {author}
  {\bibfnamefont {V.}~\bibnamefont {Scarani}},\ and\ \bibinfo {author}
  {\bibfnamefont {G.~T.}\ \bibnamefont {Landi}},\ }\href
  {https://doi.org/10.1103/PhysRevLett.123.180602} {\bibfield  {journal}
  {\bibinfo  {journal} {Physical Review Letters}\ }\textbf {\bibinfo {volume}
  {123}},\ \bibinfo {pages} {180602} (\bibinfo {year}
  {2019}{\natexlab{b}})}\BibitemShut {NoStop}%
\bibitem [{\citenamefont {De~Chiara}\ and\ \citenamefont
  {Antezza}(2020)}]{de_chiara_quantum_2020}%
  \BibitemOpen
  \bibfield  {author} {\bibinfo {author} {\bibfnamefont {G.}~\bibnamefont
  {De~Chiara}}\ and\ \bibinfo {author} {\bibfnamefont {M.}~\bibnamefont
  {Antezza}},\ }\href {https://doi.org/10.1103/PhysRevResearch.2.033315}
  {\bibfield  {journal} {\bibinfo  {journal} {Physical Review Research}\
  }\textbf {\bibinfo {volume} {2}},\ \bibinfo {pages} {033315} (\bibinfo {year}
  {2020})}\BibitemShut {NoStop}%
\bibitem [{\citenamefont {Strasberg}(2019)}]{strasberg_repeated_2019}%
  \BibitemOpen
  \bibfield  {author} {\bibinfo {author} {\bibfnamefont {P.}~\bibnamefont
  {Strasberg}},\ }\href {https://doi.org/10.1103/PhysRevLett.123.180604}
  {\bibfield  {journal} {\bibinfo  {journal} {Physical Review Letters}\
  }\textbf {\bibinfo {volume} {123}},\ \bibinfo {pages} {180604} (\bibinfo
  {year} {2019})}\BibitemShut {NoStop}%
\bibitem [{\citenamefont {Minsky}\ and\ \citenamefont
  {Papert}(1987)}]{minsky_perceptrons_1987}%
  \BibitemOpen
  \bibfield  {author} {\bibinfo {author} {\bibfnamefont {M.}~\bibnamefont
  {Minsky}}\ and\ \bibinfo {author} {\bibfnamefont {S.~A.}\ \bibnamefont
  {Papert}},\ }\href@noop {} {\emph {\bibinfo {title} {Perceptrons: {An}
  {Introduction} to {Computational} {Geometry}, {Expanded} {Edition}}}},\
  \bibinfo {edition} {expanded, subsequent edition}\ ed.\ (\bibinfo
  {publisher} {The M.I.T. Press},\ \bibinfo {address} {Cambridge, Mass},\
  \bibinfo {year} {1987})\BibitemShut {NoStop}%
\bibitem [{\citenamefont {Apicella}\ \emph {et~al.}(2021)\citenamefont
  {Apicella}, \citenamefont {Donnarumma}, \citenamefont {Isgrò},\ and\
  \citenamefont {Prevete}}]{apicella_survey_2021}%
  \BibitemOpen
  \bibfield  {author} {\bibinfo {author} {\bibfnamefont {A.}~\bibnamefont
  {Apicella}}, \bibinfo {author} {\bibfnamefont {F.}~\bibnamefont
  {Donnarumma}}, \bibinfo {author} {\bibfnamefont {F.}~\bibnamefont {Isgrò}},\
  and\ \bibinfo {author} {\bibfnamefont {R.}~\bibnamefont {Prevete}},\ }\href
  {https://doi.org/10.1016/j.neunet.2021.01.026} {\bibfield  {journal}
  {\bibinfo  {journal} {Neural Networks}\ }\textbf {\bibinfo {volume} {138}},\
  \bibinfo {pages} {14} (\bibinfo {year} {2021})}\BibitemShut {NoStop}%
\bibitem [{\citenamefont {Türkpençe}(2020)}]{turkpence_reservoir_2020}%
  \BibitemOpen
  \bibfield  {author} {\bibinfo {author} {\bibfnamefont {D.}~\bibnamefont
  {Türkpençe}},\ }\href {https://doi.org/10.1016/j.physleta.2020.126442}
  {\bibfield  {journal} {\bibinfo  {journal} {Physics Letters A}\ }\textbf
  {\bibinfo {volume} {384}},\ \bibinfo {pages} {126442} (\bibinfo {year}
  {2020})}\BibitemShut {NoStop}%
\bibitem [{\citenamefont {Filippov}\ \emph {et~al.}(2017)\citenamefont
  {Filippov}, \citenamefont {Piilo}, \citenamefont {Maniscalco},\ and\
  \citenamefont {Ziman}}]{filippov_divisibility_2017}%
  \BibitemOpen
  \bibfield  {author} {\bibinfo {author} {\bibfnamefont {S.~N.}\ \bibnamefont
  {Filippov}}, \bibinfo {author} {\bibfnamefont {J.}~\bibnamefont {Piilo}},
  \bibinfo {author} {\bibfnamefont {S.}~\bibnamefont {Maniscalco}},\ and\
  \bibinfo {author} {\bibfnamefont {M.}~\bibnamefont {Ziman}},\ }\href
  {https://doi.org/10.1103/PhysRevA.96.032111} {\bibfield  {journal} {\bibinfo
  {journal} {Physical Review A}\ }\textbf {\bibinfo {volume} {96}},\ \bibinfo
  {pages} {032111} (\bibinfo {year} {2017})}\BibitemShut {NoStop}%
\bibitem [{\citenamefont {Kołodyński}\ \emph {et~al.}(2018)\citenamefont
  {Kołodyński}, \citenamefont {Brask}, \citenamefont {Perarnau-Llobet},\ and\
  \citenamefont {Bylicka}}]{kolodynski_adding_2018}%
  \BibitemOpen
  \bibfield  {author} {\bibinfo {author} {\bibfnamefont {J.}~\bibnamefont
  {Kołodyński}}, \bibinfo {author} {\bibfnamefont {J.~B.}\ \bibnamefont
  {Brask}}, \bibinfo {author} {\bibfnamefont {M.}~\bibnamefont
  {Perarnau-Llobet}},\ and\ \bibinfo {author} {\bibfnamefont {B.}~\bibnamefont
  {Bylicka}},\ }\href {https://doi.org/10.1103/PhysRevA.97.062124} {\bibfield
  {journal} {\bibinfo  {journal} {Physical Review A}\ }\textbf {\bibinfo
  {volume} {97}},\ \bibinfo {pages} {062124} (\bibinfo {year}
  {2018})}\BibitemShut {NoStop}%
\bibitem [{\citenamefont {Mitchison}\ and\ \citenamefont
  {Plenio}(2018)}]{mitchison_non-additive_2018}%
  \BibitemOpen
  \bibfield  {author} {\bibinfo {author} {\bibfnamefont {M.~T.}\ \bibnamefont
  {Mitchison}}\ and\ \bibinfo {author} {\bibfnamefont {M.~B.}\ \bibnamefont
  {Plenio}},\ }\href {https://doi.org/10.1088/1367-2630/aa9f70} {\bibfield
  {journal} {\bibinfo  {journal} {New Journal of Physics}\ }\textbf {\bibinfo
  {volume} {20}},\ \bibinfo {pages} {033005} (\bibinfo {year}
  {2018})}\BibitemShut {NoStop}%
\bibitem [{\citenamefont {Helstrom}(1969)}]{helstrom_quantum_1969}%
  \BibitemOpen
  \bibfield  {author} {\bibinfo {author} {\bibfnamefont {C.~W.}\ \bibnamefont
  {Helstrom}},\ }\href {https://doi.org/10.1007/BF01007479} {\bibfield
  {journal} {\bibinfo  {journal} {Journal of Statistical Physics}\ }\textbf
  {\bibinfo {volume} {1}},\ \bibinfo {pages} {231} (\bibinfo {year}
  {1969})}\BibitemShut {NoStop}%
\bibitem [{\citenamefont {Dittmann}(1999)}]{dittmann_explicit_1999}%
  \BibitemOpen
  \bibfield  {author} {\bibinfo {author} {\bibfnamefont {J.}~\bibnamefont
  {Dittmann}},\ }\href {https://doi.org/10.1088/0305-4470/32/14/007} {\bibfield
   {journal} {\bibinfo  {journal} {Journal of Physics A: Mathematical and
  General}\ }\textbf {\bibinfo {volume} {32}},\ \bibinfo {pages} {2663}
  (\bibinfo {year} {1999})}\BibitemShut {NoStop}%
\bibitem [{\citenamefont {Zhong}\ \emph {et~al.}(2013)\citenamefont {Zhong},
  \citenamefont {Sun}, \citenamefont {Ma}, \citenamefont {Wang},\ and\
  \citenamefont {Nori}}]{zhong_fisher_2013}%
  \BibitemOpen
  \bibfield  {author} {\bibinfo {author} {\bibfnamefont {W.}~\bibnamefont
  {Zhong}}, \bibinfo {author} {\bibfnamefont {Z.}~\bibnamefont {Sun}}, \bibinfo
  {author} {\bibfnamefont {J.}~\bibnamefont {Ma}}, \bibinfo {author}
  {\bibfnamefont {X.}~\bibnamefont {Wang}},\ and\ \bibinfo {author}
  {\bibfnamefont {F.}~\bibnamefont {Nori}},\ }\href
  {https://doi.org/10.1103/PhysRevA.87.022337} {\bibfield  {journal} {\bibinfo
  {journal} {Physical Review A}\ }\textbf {\bibinfo {volume} {87}},\ \bibinfo
  {pages} {022337} (\bibinfo {year} {2013})}\BibitemShut {NoStop}%
\bibitem [{\citenamefont {Ziman}\ and\ \citenamefont
  {Bužek}(2005)}]{ziman_all_2005}%
  \BibitemOpen
  \bibfield  {author} {\bibinfo {author} {\bibfnamefont {M.}~\bibnamefont
  {Ziman}}\ and\ \bibinfo {author} {\bibfnamefont {V.}~\bibnamefont {Bužek}},\
  }\href {https://doi.org/10.1103/PhysRevA.72.022110} {\bibfield  {journal}
  {\bibinfo  {journal} {Physical Review A}\ }\textbf {\bibinfo {volume} {72}},\
  \bibinfo {pages} {022110} (\bibinfo {year} {2005})},\ \bibinfo {note}
  {publisher: American Physical Society}\BibitemShut {NoStop}%
\bibitem [{\citenamefont {Cresser}(1992)}]{cresser_quantum-field_1992}%
  \BibitemOpen
  \bibfield  {author} {\bibinfo {author} {\bibfnamefont {J.~D.}\ \bibnamefont
  {Cresser}},\ }\href {https://doi.org/10.1103/PhysRevA.46.5913} {\bibfield
  {journal} {\bibinfo  {journal} {Physical Review A}\ }\textbf {\bibinfo
  {volume} {46}},\ \bibinfo {pages} {5913} (\bibinfo {year}
  {1992})}\BibitemShut {NoStop}%
\bibitem [{\citenamefont {Liao}\ \emph {et~al.}(2010)\citenamefont {Liao},
  \citenamefont {Dong},\ and\ \citenamefont {Sun}}]{liao_single-particle_2010}%
  \BibitemOpen
  \bibfield  {author} {\bibinfo {author} {\bibfnamefont {J.-Q.}\ \bibnamefont
  {Liao}}, \bibinfo {author} {\bibfnamefont {H.}~\bibnamefont {Dong}},\ and\
  \bibinfo {author} {\bibfnamefont {C.~P.}\ \bibnamefont {Sun}},\ }\href
  {https://doi.org/10.1103/PhysRevA.81.052121} {\bibfield  {journal} {\bibinfo
  {journal} {Physical Review A}\ }\textbf {\bibinfo {volume} {81}},\ \bibinfo
  {pages} {052121} (\bibinfo {year} {2010})}\BibitemShut {NoStop}%
\bibitem [{\citenamefont {Türkpençe}\ and\ \citenamefont
  {M\"{u}stecapl{\i}o\u{g}lu}(2016)}]{turkpence_quantum_2016}%
  \BibitemOpen
  \bibfield  {author} {\bibinfo {author} {\bibfnamefont {D.}~\bibnamefont
  {Türkpençe}}\ and\ \bibinfo {author} {\bibfnamefont {\"{O}.~E.}\ \bibnamefont
  {M\"{u}stecapl{\i}o\u{g}lu}},\ }\href {https://doi.org/10.1103/PhysRevE.93.012145}
  {\bibfield  {journal} {\bibinfo  {journal} {Physical Review E}\ }\textbf
  {\bibinfo {volume} {93}},\ \bibinfo {pages} {012145} (\bibinfo {year}
  {2016})}\BibitemShut {NoStop}%
\bibitem [{\citenamefont {Filipowicz}\ \emph {et~al.}(1986)\citenamefont
  {Filipowicz}, \citenamefont {Javanainen},\ and\ \citenamefont
  {Meystre}}]{filipowicz_theory_1986}%
  \BibitemOpen
  \bibfield  {author} {\bibinfo {author} {\bibfnamefont {P.}~\bibnamefont
  {Filipowicz}}, \bibinfo {author} {\bibfnamefont {J.}~\bibnamefont
  {Javanainen}},\ and\ \bibinfo {author} {\bibfnamefont {P.}~\bibnamefont
  {Meystre}},\ }\href {https://doi.org/10.1103/PhysRevA.34.3077} {\bibfield
  {journal} {\bibinfo  {journal} {Physical Review A}\ }\textbf {\bibinfo
  {volume} {34}},\ \bibinfo {pages} {3077} (\bibinfo {year}
  {1986})}\BibitemShut {NoStop}%
\bibitem [{\citenamefont {Korkmaz}\ and\ \citenamefont
  {Türkpençe}(2022)}]{korkmaz_transfer_2022}%
  \BibitemOpen
  \bibfield  {author} {\bibinfo {author} {\bibfnamefont {U.}~\bibnamefont
  {Korkmaz}}\ and\ \bibinfo {author} {\bibfnamefont {D.}~\bibnamefont
  {Türkpençe}},\ }\href {https://doi.org/10.1016/j.physleta.2021.127887}
  {\bibfield  {journal} {\bibinfo  {journal} {Physics Letters A}\ }\textbf
  {\bibinfo {volume} {426}},\ \bibinfo {pages} {127887} (\bibinfo {year}
  {2022})}\BibitemShut {NoStop}%
\bibitem [{\citenamefont {Johansson}\ \emph {et~al.}(2013)\citenamefont
  {Johansson}, \citenamefont {Nation},\ and\ \citenamefont
  {Nori}}]{johansson_qutip_2013}%
  \BibitemOpen
  \bibfield  {author} {\bibinfo {author} {\bibfnamefont {J.~R.}\ \bibnamefont
  {Johansson}}, \bibinfo {author} {\bibfnamefont {P.~D.}\ \bibnamefont
  {Nation}},\ and\ \bibinfo {author} {\bibfnamefont {F.}~\bibnamefont {Nori}},\
  }\href {https://doi.org/10.1016/j.cpc.2012.11.019} {\bibfield  {journal}
  {\bibinfo  {journal} {Computer Physics Communications}\ }\textbf {\bibinfo
  {volume} {184}},\ \bibinfo {pages} {1234} (\bibinfo {year}
  {2013})}\BibitemShut {NoStop}%
\bibitem [{\citenamefont {Blais}\ \emph {et~al.}(2021)\citenamefont {Blais},
  \citenamefont {Grimsmo}, \citenamefont {Girvin},\ and\ \citenamefont
  {Wallraff}}]{blais_circuit_2021}%
  \BibitemOpen
  \bibfield  {author} {\bibinfo {author} {\bibfnamefont {A.}~\bibnamefont
  {Blais}}, \bibinfo {author} {\bibfnamefont {A.~L.}\ \bibnamefont {Grimsmo}},
  \bibinfo {author} {\bibfnamefont {S.}~\bibnamefont {Girvin}},\ and\ \bibinfo
  {author} {\bibfnamefont {A.}~\bibnamefont {Wallraff}},\ }\href
  {https://doi.org/10.1103/RevModPhys.93.025005} {\bibfield  {journal}
  {\bibinfo  {journal} {Reviews of Modern Physics}\ }\textbf {\bibinfo {volume}
  {93}},\ \bibinfo {pages} {025005} (\bibinfo {year} {2021})}\BibitemShut
  {NoStop}%
\bibitem [{\citenamefont {Krantz}\ \emph {et~al.}(2019)\citenamefont {Krantz},
  \citenamefont {Kjaergaard}, \citenamefont {Yan}, \citenamefont {Orlando},
  \citenamefont {Gustavsson},\ and\ \citenamefont
  {Oliver}}]{krantz_quantum_2019}%
  \BibitemOpen
  \bibfield  {author} {\bibinfo {author} {\bibfnamefont {P.}~\bibnamefont
  {Krantz}}, \bibinfo {author} {\bibfnamefont {M.}~\bibnamefont {Kjaergaard}},
  \bibinfo {author} {\bibfnamefont {F.}~\bibnamefont {Yan}}, \bibinfo {author}
  {\bibfnamefont {T.~P.}\ \bibnamefont {Orlando}}, \bibinfo {author}
  {\bibfnamefont {S.}~\bibnamefont {Gustavsson}},\ and\ \bibinfo {author}
  {\bibfnamefont {W.~D.}\ \bibnamefont {Oliver}},\ }\href
  {https://doi.org/10.1063/1.5089550} {\bibfield  {journal} {\bibinfo
  {journal} {Applied Physics Reviews}\ }\textbf {\bibinfo {volume} {6}},\
  \bibinfo {pages} {021318} (\bibinfo {year} {2019})}\BibitemShut {NoStop}%
\bibitem [{\citenamefont {Deng}\ \emph {et~al.}(2017)\citenamefont {Deng},
  \citenamefont {Barnes},\ and\ \citenamefont
  {Economou}}]{deng_robustness_2017}%
  \BibitemOpen
  \bibfield  {author} {\bibinfo {author} {\bibfnamefont {X.-H.}\ \bibnamefont
  {Deng}}, \bibinfo {author} {\bibfnamefont {E.}~\bibnamefont {Barnes}},\ and\
  \bibinfo {author} {\bibfnamefont {S.~E.}\ \bibnamefont {Economou}},\ }\href
  {https://doi.org/10.1103/PhysRevB.96.035441} {\bibfield  {journal} {\bibinfo
  {journal} {Physical Review B}\ }\textbf {\bibinfo {volume} {96}},\ \bibinfo
  {pages} {035441} (\bibinfo {year} {2017})}\BibitemShut {NoStop}%
\bibitem [{\citenamefont {Majer}\ \emph {et~al.}(2007)\citenamefont {Majer},
  \citenamefont {Chow}, \citenamefont {Gambetta}, \citenamefont {Koch},
  \citenamefont {Johnson}, \citenamefont {Schreier}, \citenamefont {Frunzio},
  \citenamefont {Schuster}, \citenamefont {Houck}, \citenamefont {Wallraff},
  \citenamefont {Blais}, \citenamefont {Devoret}, \citenamefont {Girvin},\ and\
  \citenamefont {Schoelkopf}}]{majer_coupling_2007}%
  \BibitemOpen
  \bibfield  {author} {\bibinfo {author} {\bibfnamefont {J.}~\bibnamefont
  {Majer}}, \bibinfo {author} {\bibfnamefont {J.~M.}\ \bibnamefont {Chow}},
  \bibinfo {author} {\bibfnamefont {J.~M.}\ \bibnamefont {Gambetta}}, \bibinfo
  {author} {\bibfnamefont {J.}~\bibnamefont {Koch}}, \bibinfo {author}
  {\bibfnamefont {B.~R.}\ \bibnamefont {Johnson}}, \bibinfo {author}
  {\bibfnamefont {J.~A.}\ \bibnamefont {Schreier}}, \bibinfo {author}
  {\bibfnamefont {L.}~\bibnamefont {Frunzio}}, \bibinfo {author} {\bibfnamefont
  {D.~I.}\ \bibnamefont {Schuster}}, \bibinfo {author} {\bibfnamefont {A.~A.}\
  \bibnamefont {Houck}}, \bibinfo {author} {\bibfnamefont {A.}~\bibnamefont
  {Wallraff}}, \bibinfo {author} {\bibfnamefont {A.}~\bibnamefont {Blais}},
  \bibinfo {author} {\bibfnamefont {M.~H.}\ \bibnamefont {Devoret}}, \bibinfo
  {author} {\bibfnamefont {S.~M.}\ \bibnamefont {Girvin}},\ and\ \bibinfo
  {author} {\bibfnamefont {R.~J.}\ \bibnamefont {Schoelkopf}},\ }\href
  {https://doi.org/10.1038/nature06184} {\bibfield  {journal} {\bibinfo
  {journal} {Nature}\ }\textbf {\bibinfo {volume} {449}},\ \bibinfo {pages}
  {443} (\bibinfo {year} {2007})}\BibitemShut {NoStop}%
\bibitem [{res(9 28)}]{research_ibm}%
  \BibitemOpen
  \href@noop {} {\bibinfo {title} {{IBM} {Q} {Experience}}},\ \bibinfo
  {howpublished} {\url{https://quantum-computing.ibm.com/}} (\bibinfo {year}
  {Accessed: 2021-09-28})\BibitemShut {NoStop}%
\bibitem [{\citenamefont {Zhou}\ \emph {et~al.}(2021)\citenamefont {Zhou},
  \citenamefont {Zhang}, \citenamefont {Yin}, \citenamefont {Huai},
  \citenamefont {Gu}, \citenamefont {Xu}, \citenamefont {Allcock},
  \citenamefont {Liu}, \citenamefont {Xi}, \citenamefont {Yu}, \citenamefont
  {Zhang}, \citenamefont {Zhang}, \citenamefont {Li}, \citenamefont {Song},
  \citenamefont {Wang}, \citenamefont {Zheng}, \citenamefont {An},
  \citenamefont {Zheng},\ and\ \citenamefont {Zhang}}]{zhou_rapid_2021}%
  \BibitemOpen
  \bibfield  {author} {\bibinfo {author} {\bibfnamefont {Y.}~\bibnamefont
  {Zhou}}, \bibinfo {author} {\bibfnamefont {Z.}~\bibnamefont {Zhang}},
  \bibinfo {author} {\bibfnamefont {Z.}~\bibnamefont {Yin}}, \bibinfo {author}
  {\bibfnamefont {S.}~\bibnamefont {Huai}}, \bibinfo {author} {\bibfnamefont
  {X.}~\bibnamefont {Gu}}, \bibinfo {author} {\bibfnamefont {X.}~\bibnamefont
  {Xu}}, \bibinfo {author} {\bibfnamefont {J.}~\bibnamefont {Allcock}},
  \bibinfo {author} {\bibfnamefont {F.}~\bibnamefont {Liu}}, \bibinfo {author}
  {\bibfnamefont {G.}~\bibnamefont {Xi}}, \bibinfo {author} {\bibfnamefont
  {Q.}~\bibnamefont {Yu}}, \bibinfo {author} {\bibfnamefont {H.}~\bibnamefont
  {Zhang}}, \bibinfo {author} {\bibfnamefont {M.}~\bibnamefont {Zhang}},
  \bibinfo {author} {\bibfnamefont {H.}~\bibnamefont {Li}}, \bibinfo {author}
  {\bibfnamefont {X.}~\bibnamefont {Song}}, \bibinfo {author} {\bibfnamefont
  {Z.}~\bibnamefont {Wang}}, \bibinfo {author} {\bibfnamefont {D.}~\bibnamefont
  {Zheng}}, \bibinfo {author} {\bibfnamefont {S.}~\bibnamefont {An}}, \bibinfo
  {author} {\bibfnamefont {Y.}~\bibnamefont {Zheng}},\ and\ \bibinfo {author}
  {\bibfnamefont {S.}~\bibnamefont {Zhang}},\ }\href@noop {} {\bibfield
  {journal} {\bibinfo  {journal} {Nature Communications}\ }\textbf {\bibinfo
  {volume} {12}},\ \bibinfo {pages} {5924} (\bibinfo {year}
  {2021})}\BibitemShut {NoStop}%
\bibitem [{\citenamefont {Tan}\ \emph {et~al.}(2022)\citenamefont {Tan},
  \citenamefont {Wu}, \citenamefont {Xu}, \citenamefont {Liu},\ and\
  \citenamefont {Kuang}}]{tan_quantum_2022}%
  \BibitemOpen
  \bibfield  {author} {\bibinfo {author} {\bibfnamefont {Q.-S.}\ \bibnamefont
  {Tan}}, \bibinfo {author} {\bibfnamefont {W.}~\bibnamefont {Wu}}, \bibinfo
  {author} {\bibfnamefont {L.}~\bibnamefont {Xu}}, \bibinfo {author}
  {\bibfnamefont {J.}~\bibnamefont {Liu}},\ and\ \bibinfo {author}
  {\bibfnamefont {L.-M.}\ \bibnamefont {Kuang}},\ }\href
  {https://doi.org/10.1103/PhysRevA.106.032602} {\bibfield  {journal} {\bibinfo
   {journal} {Physical Review A}\ }\textbf {\bibinfo {volume} {106}},\ \bibinfo
  {pages} {032602} (\bibinfo {year} {2022})}\BibitemShut {NoStop}%
\bibitem [{\citenamefont {Sha}\ and\ \citenamefont
  {Wu}(2022)}]{sha_continuous-variable_2022}%
  \BibitemOpen
  \bibfield  {author} {\bibinfo {author} {\bibfnamefont {Y.-D.}\ \bibnamefont
  {Sha}}\ and\ \bibinfo {author} {\bibfnamefont {W.}~\bibnamefont {Wu}},\
  }\href {https://doi.org/10.1103/PhysRevResearch.4.023169} {\bibfield
  {journal} {\bibinfo  {journal} {Physical Review Research}\ }\textbf {\bibinfo
  {volume} {4}},\ \bibinfo {pages} {023169} (\bibinfo {year}
  {2022})}\BibitemShut {NoStop}%
\bibitem [{\citenamefont {Wu}\ \emph {et~al.}(2021)\citenamefont {Wu},
  \citenamefont {Bai},\ and\ \citenamefont {An}}]{wu_non-markovian_2021}%
  \BibitemOpen
  \bibfield  {author} {\bibinfo {author} {\bibfnamefont {W.}~\bibnamefont
  {Wu}}, \bibinfo {author} {\bibfnamefont {S.-Y.}\ \bibnamefont {Bai}},\ and\
  \bibinfo {author} {\bibfnamefont {J.-H.}\ \bibnamefont {An}},\ }\href
  {https://doi.org/10.1103/PhysRevA.103.L010601} {\bibfield  {journal}
  {\bibinfo  {journal} {Physical Review A}\ }\textbf {\bibinfo {volume}
  {103}},\ \bibinfo {pages} {L010601} (\bibinfo {year} {2021})}\BibitemShut
  {NoStop}%
\end{thebibliography}
%

\end{document}